\documentclass[preprint]{revtex4-1}

\usepackage{amssymb,amsfonts,amsmath}
\usepackage{graphicx}
\usepackage{color}
\usepackage{url,hyperref,lineno,microtype,subcaption}
\usepackage{mathtools}

\newcommand\timerev{\stackrel{\mathclap{\normalfont\mbox{t.r.}}}{\leftrightarrow}}

\begin{document}

\title{Time irreversibility of resting brain activity in the healthy brain and pathology} 

\author{Massimiliano Zanin} 
\affiliation{Centro de Tecnolog\'ia Biom\'edica, Universidad Polit\'ecnica de Madrid, Madrid, Spain} 
\email{massimiliano.zanin@gmail.com} 

\author{Bahar G\"untekin} 
\affiliation{Department of Biophysics, International School of Medicine, Istanbul Medipol University, Istanbul, Turkey} 
\affiliation{REMER, Clinical Electrophysiology, Neuroimaging and Neuromodulation Lab, Istanbul Medipol University, Istanbul, Turkey}

\author{Tuba Akt\"urk}
\affiliation{REMER, Clinical Electrophysiology, Neuroimaging and Neuromodulation Lab, Istanbul Medipol University, Istanbul, Turkey}
\affiliation{Istanbul Medipol University, Vocational School, Program of Electroneurophysiology, Istanbul, Turkey}

\author{L\"utf\"u Hano\u{g}lu}
\affiliation{REMER, Clinical Electrophysiology, Neuroimaging and Neuromodulation Lab, Istanbul Medipol University, Istanbul, Turkey}
\affiliation{Istanbul Medipol University, School of Medicine, Department of Neurology, Istanbul, Turkey}

\author{David Papo} 
\affiliation{SCALab, CNRS, Universit\'e de Lille, Villeneuve d'Ascq, France}

\begin{abstract}

Characterising brain activity at rest is of paramount importance to our understanding both of general principles of brain functioning and of the way brain dynamics is affected in the presence of neurological or psychiatric pathologies. We measured the time-reversal symmetry of spontaneous electroencephalographic brain activity recorded from three groups of patients and their respective control group under two experimental conditions (eyes open and closed). We evaluated differences in time irreversibility in terms of possible underlying physical generating mechanisms. The results showed that resting brain activity is generically time-irreversible at sufficiently long time scales, and that brain pathology is generally associated with a reduction in time-asymmetry, albeit with pathology-specific patterns. The significance of these results and their possible dynamical aetiology are discussed. Some implications of the differential modulation of time asymmetry by pathology and experimental condition are examined.
\end{abstract}


\maketitle

\section{Introduction}

Even in the absence of exogenous stimulation and for constant values of the parameters controlling its dynamics, the brain generates fluctuations characterised by non-random patterns over a wide range of spatial and temporal scales \citep{arieli1996dynamics, van2010eeg, deco2011emerging} re-edited across the cortical space in a non-random way \citep{kenet2003spontaneously, cossart2003attractor, beggs2004neuronal, ikegaya2004synfire, dragoi2011preplay, betzel2012synchronization}.

Characterising resting activity is important for at least three main partially interrelated reasons. On the one hand, accumulating evidence shows that neurological and psychiatric conditions are associated with alterations of several aspects of resting activity structure \citep{zhang2010disease, alderson2015hearing, hohenfeld2018resting}. On the other hand, spontaneous fluctuations are intimately related to stimulus-induced ones \citep{shew2009neuronal, luczak2009spontaneous, smith2009correspondence}, so that characterising the former also provides insight onto the latter. No less importantly, the structure of resting brain activity fluctuations gives away key aspects of the physics of the underlying system producing them \citep{papo2014functional}. For instance, if the brain is understood as a complex thermodynamic machine, the activity recorded with standard system-level neuroimaging techniques can be thought of as thermal fluctuations through which the energy is dissipated to ensure its functioning \citep{livi2013brain}. Within this framework, the generic complex spatio-temporal scaling properties of resting brain activity, including scale invariance and long-range temporal memory \citep{novikov1997scale, linkenkaer2001long, gong2007intermittent, bianco2007brain, freyer2009bistability, expert2011self, wink2008monofractal}, can be understood as indicators of the fact that the brain operates away from equilibrium \citep{papo2013should}. 

Quantifying the extent to which a system such as the brain deviates from equilibrium conditions is an important issue. The fluctuations of a system at equilibrium obey detailed balance of the probability fluxes, a condition whereby the net current between any pair of states vanishes at long enough times, i.e. given two states $x$ and $y$ and a transition rate $W(\cdot)$ following condition holds: $\rho(x)W(x \rightarrow y) = \rho(y)W(y \rightarrow x)$, where $\rho(\cdot)$ is the equilibrium probability distribution. Importantly, this condition can be understood in terms of symmetry property of the probability distributions $P(\omega_t) = P(I\omega_t)$ of a trajectory $\omega_t = (\omega_1, \omega_2, \ldots, \omega_t)$ of length $t$ and its time-reversed one, where $I$ denotes the time reverse operator. In systems outside of equilibrium, this symmetry is broken due to the presence of non-conservative forces: energy dissipation happens with an irreversible increase of entropy, and the time reversal symmetry is then broken. Beyond such explicit dissipation, irreversibility can also be due to the presence of memory, which acts as a hidden dissipative external force in a process \citep{puglisi2009irreversible}; and, it is destroyed by the presence of noise \citep{porporato2007irreversibility, xia2014classifying}. 

Time irreversibility provides valuable information on the statistical properties of the generating processes of given stochastic dynamics. On the one hand, reversibility implies stationarity \citep{lawrance1991directionality}. On the other hand, linear Gaussian random processes and static nonlinear transformations of such processes are reversible, and significant time irreversibility excludes Gaussian linear processes or linear ARMA models as possible generating dynamics, implying instead nonlinear dynamics, (linear or nonlinear) non-Gaussian, \citep{weiss1975time, cox1981statistical, lawrance1991directionality, stone1996detecting}. The asymmetry under time reversal of some system variable's statistical properties provides a quantitative estimate of the thermodynamic {\it entropy production} $\Sigma_t$ of the system generating the activity, even when the details of the system are unknown \citep{gaspard2005brownian, andrieux2007entropy, roldan2010estimating}. Note that the coarse-grained entropy production provides a lower bound on the true one \citep{seifert2019stochastic}.
This fundamental relation between thermodynamic entropy (a macroscopic quantity) and Kolmogorov-Sinai entropy (a microscopic quantity) has in particular been proven to hold for systems in non-equilibrium steady state (NESS) \citep{gaspard2004time, roldan2010estimating}. $\Sigma_t$ can be represented in terms of the ratio $\Sigma_t = ln P(\omega_t) / P(I \omega_t)$. This quantity is identically equal to zero for each trajectory separately if detailed balance is satisfied, but always non-negative otherwise. Non-equilibrium systems obey {\it fluctuation relations} which hold for any stationary time series, independently of their dynamics \citep{evans1993probability, gallavotti1995dynamical, crooks2000path, evans2002fluctuation}. In particular, the following relation 

\begin{equation}
	P(-\Sigma_t) \sim P(\Sigma_t) e^{-\Sigma_t}
	\label{eq:P}
\end{equation}

provides a quantitative expression for the probability of entropy of a finite non-equilibrium flowing in a direction opposite to that dictated by the second law of thermodynamics, when considered in a finite time. This relation illustrates the fact that for out-equilibrium dynamics the negative tail of the probability distribution decays faster than the positive one. 

Not surprisingly, time irreversibility metrics have extensively been used to characterise real-world systems, with a special attention being devoted to economic and financial time series \cite{ramsey1996time, zumbach2009time, xia2014classifying}. 
Time reversal asymmetry has also been used to characterise healthy and pathological activity of biological systems, particularly the human heart \citep{costa2005broken, guzik2006heart, piskorski2007geometry, porta2006time, porta2008temporal, porta2009assessment, karmakar2009defining, hou2010analysis}, but also to classify hand tremor \citep{timmer1993characteristics}. However, the time-reversal symmetry properties of brain activity have attracted little attention \citep{paluvs1996nonlinearity, van1996time, ehlers1998low, visnovcova2014complexity, yao2018permutation} and have not yet been systematically examined. For instance, \cite{paluvs1996nonlinearity} found the mutual information between EEG time series and their lagged versions to be time-asymmetric. However, since the asymmetry in the peaks of the mutual information, itself symmetric, may not be equivalent to the temporal asymmetry of the underlying process, the observed properties were tentatively explained as reflecting non-stationary nonlinear deterministic oscillatory episodes randomly distributed in a noisy background. Three studies examined time irreversibility in epilepsy, consistently reporting increased irreversibility for ictal activity in both scalp and intracranial recorded electrical brain activity \citep{van1996time, schindler2016ictal, martinez2018detection}. The surgical removal of brain areas generating time-irreversible iEEG signals was associated with seizure free post-surgical outcome \citep{schindler2016ictal}.

Here we address the following main questions: what's the typical time asymmetry of brain activity at rest? How is it modified by a simple experimental condition such as opening and closing eyes? How does it vary in neurological and psychiatric brain pathologies? We conjectured that, insofar as entropy production determines the performance of thermal machines such as the brain, and disease is associated with impaired self-organising capabilities, abnormal time reversal symmetry properties may be a marker of pathology and may be differentially affected by different neurological and psychiatric diseases.
These questions are addressed by analysing a large set of EEG recordings, comprising three groups of patients and the corresponding control groups, through a recently proposed irreversibility metric based on the assessment of permutation patterns \cite{zanin2018assessing}. 
Results suggest that the human brain is generically time-irreversible; that such property is increased in eyes open resting states, with respect to eyes closed ones; and that pathologies like Parkinson's disease and schizophrenia decrease the irreversibility. We further show that irreversibility is non-trivially modified by filtering the EEG signal at different bands, and that its nature can be studied by resorting to surrogate time series.

\section{Materials and Methods}

\subsection{Assessing irreversibility in time series}
\label{sec:irreversibility}

In general terms, the time asymmetry of a stationary driven system can be determined by the Kullback-Leibler (KL) distance between probability distributions representing the forward and reverse trajectory (respectively, $p$ and $\hat p$):

\begin{equation}
	KL(p || \hat p) = \sum p(\omega_1, \omega_2, \ldots, \omega_n) \log \frac{p(\omega_1, \omega_2, \ldots, \omega_n)}{\hat p(\omega_1, \omega_2, \ldots, \omega_n)}.
	\label{eq:KLIn}
\end{equation}

The KL distance can be thought of as the mean of the difference between $p$ and $\hat p$, and quantifies the distinguishability or, loosely, the distance between these two probability distributions \citep{gaspard2005brownian, andrieux2007entropy, porporato2007irreversibility}. The KL distance is not just an estimator of entropy production's lower bound but it also provides a general method to distinguish between equilibrium and NESS \citep{roldan2010estimating}. 

While Eq. \ref{eq:KLIn} defines a general rule for estimating irreversibility, it does not define what $p$ and $\hat p$ should represent. Consequently, various methods to quantify time reversibility from empirical time series have been proposed and applied to real-world problems, particularly biological and financial systems \citep{diks1995reversibility, paluvs1996nonlinearity, ramsey1996time, daw2000symbolic, kennel2004testing, costa2005broken, costa2008multiscale, casali2008multiple, zumbach2009time, donges2013testing, xia2014classifying, lacasa2012time, lacasa2015time, flanagan2016irreversibility}. Here, we use a method \citep{graff2013ordinal, zanin2018assessing, martinez2018detection} based on permutation entropy \citep{bandt2002permutation, zanin2012permutation}. This method presents various advantages: it has no free parameters other than the embedding dimension of the permutation entropy; as visibility graph methods \citep{lacasa2012time} it is not an all-or-none measure of irreversibility, so that its use is also meaningful for nonstationary signals, which are by definition irreversible, and is temporally local, and therefore allows assessing fluctuations; however, unlike visibility graphs, it does not rely on scaling arguments and its convergence speed is faster and hypothesis testing more straightforward. For the sake of completeness, we here review the method, starting by the definition of the permutation patterns.

\subsubsection{Permutation patterns}

The idea of analysing a time series through its permutation patterns was introduced by Bandt and Pompe \citep{bandt2002permutation}, and since then received an increasing attention from the scientific community \citep{zanin2012permutation}. Given a time series $X = \{x_t\}$, with $t = 1 \ldots N$, this is usually divided in overlapping regions of length $D$, such that:

\begin{equation}
s \rightarrow (x_s, x_{s+\tau}, \ldots , x_{s+\tau(D-2)}, x_{s+\tau(D-1)} ).
\end{equation}

$D$ is called the {\it embedding dimension} and controls the quantity of information included in each region, while $\tau$ is the embedding delay. $s$ further controls the beginning of each region, and thus the degree of overlap between regions. In this study we consider $D = 3$ and $\tau = 1$.

Once these regions have been defined, an ordinal pattern is associated to each one of them. The elements composing each region are sorted in increasing order, and the ordinal pattern corresponding to the required permutation is saved for further analysis. In other words, the permutation $\pi = (r_0, r_1, \ldots, r_{D-1} )$ of $(0,1,\ldots,D-1)$ is the one fulfilling:

\begin{equation}
x_{s+r_0} \leq x_{s+r_1} \leq \ldots \leq x_{s+r_{D-2}} \leq x_{s+r_{D-1}}.
\end{equation}

\subsubsection{From permutation patterns to irreversibility}

The irreversibility of a time series is then estimated by looking at asymmetries in the appearance frequencies of the corresponding permutation patterns. Specifically, for $D = 3$, $6$ patterns can appear, paired as follows: 

\begin{eqnarray}
(0, 1, 2) \timerev (2, 1, 0) \\
(1, 0, 2) \timerev (2, 0, 1) \\
(1, 2, 0) \timerev (0, 2, 1),
\end{eqnarray}

with $\timerev$ representing a time reversal transformation. In other words, a region corresponding to the pattern $(0, 1, 2)$ (for instance, a monotonically increasing series) will become $(2, 1, 0)$ after a time reversal operation (in the previous example, it will become a monotonically decreasing series). A time series will thus be reversible if and only if all permutation patterns composing the previous pairs appear with approximatively the same frequency; if this does not happen, a time arrow can be derived from the predominant presence of one of the patterns composing the pair. In other words, and to illustrate, suppose a trivially irreversible time series with monotonically increasing values; only one permutation pattern can appear, i.e. $(0, 1, 2)$, which will transform to $(2, 1, 0)$ under a time reversal transformation. Given a new realisation of the same time series, assessing the relative abundance of $(0, 1, 2)$ over $(2, 1, 0)$ will allow to easily define if we are looking at the original or at the time reversed time series. This is nevertheless not possible is the appearance probabilities of both patterns is approximately the same.

A statistical test can easily be designed, by comparing the probability distributions of patterns in the forward and reversed time series. Specifically, if the time series is reversible, the number of times the two permutation patterns forming a pair appear should be similar - i.e. should not be different, in a statistical sense. Following the previous example, let us denote by $n_{(0, 1, 2)}$ and $n_{(2, 1, 0)}$ respectively the number of times the patterns $(0, 1, 2)$ and $(2, 1, 0)$ have appeared; and let us define:

\begin{equation}
p = \frac{ n_{(0, 1, 2)} }{ n_{(0, 1, 2)} + n_{(2, 1, 0)} }.
\end{equation}

The time series is not reversible if we can reject the null hypothesis that $p = 0.5$ in a two-sided binomial test. Note that the test should be repeated for all pairs of permutation patterns - three times in the case of $D = 3$.

\subsubsection{Representing the irreversibility of EEG data}
\label{sec:representing}

The previously described test yields a result that could {\it prima facie} be used to understand brain dynamics, i.e. one could simply assess whether or not an EEG time series is irreversible. This direct approach nevertheless masks important information, as it tells nothing about the time scales at which such irreversibility appears; may be sensitive to noise; and could be misleading when comparing time series of different lengths, as one could not exclude that the non-irreversibility of a short time series may be due to its reduced length, and not to a reversible underlying dynamics.

We here solve this problem by calculating how the irreversibility evolves as a function of the scale over which such irreversibility is assessed. To illustrate, let us consider an EEG time series composed of $N$ data points, and a window length (the irreversibility scale) of $n$, such that $n < N$. We firstly extract all overlapping sub-regions of size $n$, and evaluate their irreversibility; if at least a $90\%$ of those sub-regions are irreversible in a statistically significant way ($\alpha = 0.01$), then the whole time series is considered as irreversible for the time scale $n$. Finally, we average over all channels and all trials / subjects of a data set, to obtain the fraction of times a channel has been detected as irreversible at a given time scale $n$, and the evolution of such fraction as a function of $n$.

\subsubsection{Model of noisy irreversible time series}
\label{sec:noisy}

In order to assess whether the irreversibility evolution may only be due to noise, we here consider a simple dynamical model contaminated with additive Gaussian noise. The chosen model is the well-known logistic map, defined as:

\begin{equation}
	x_{t+1} = r x_n (1 - x_n) + \sigma \xi.
\end{equation}

$r$ is a parameter defining the dynamics of the map, here fixed to $4$ to ensure a chaotic evolution. Additionally, $\sigma$ is a parameter defining the quantity of additive noise, and $\xi$ random numbers drawn from a normal distribution $\mathcal{N}(0, 1)$.

The logistic map has here been chosen as it presents a non-trivial dynamics, but at the same time its irreversibility can be detected even in short time series \cite{zanin2018assessing}.

\subsubsection{Testing irreversibility through surrogate time series}
\label{sec:surrogate}

As a final issue, we further analyse the source of brain irreversibility by using surrogate time series - see Section \ref{sec:nature}. Such series are obtained through the Iterative Amplitude Adjusted Fourier Transform (IAAFT) algorithm \citep{schreiber1996improved}. IAAFT works by iteratively performing a random phase transformation of the original time series, aimed at creating surrogates that preserve both the linear (auto-)correlation and the amplitude of the signal.

\subsection{EEG data sets}
\label{sec:datasets}

\begin{figure*}[!tb]
\centering
\includegraphics[width=0.9\textwidth]{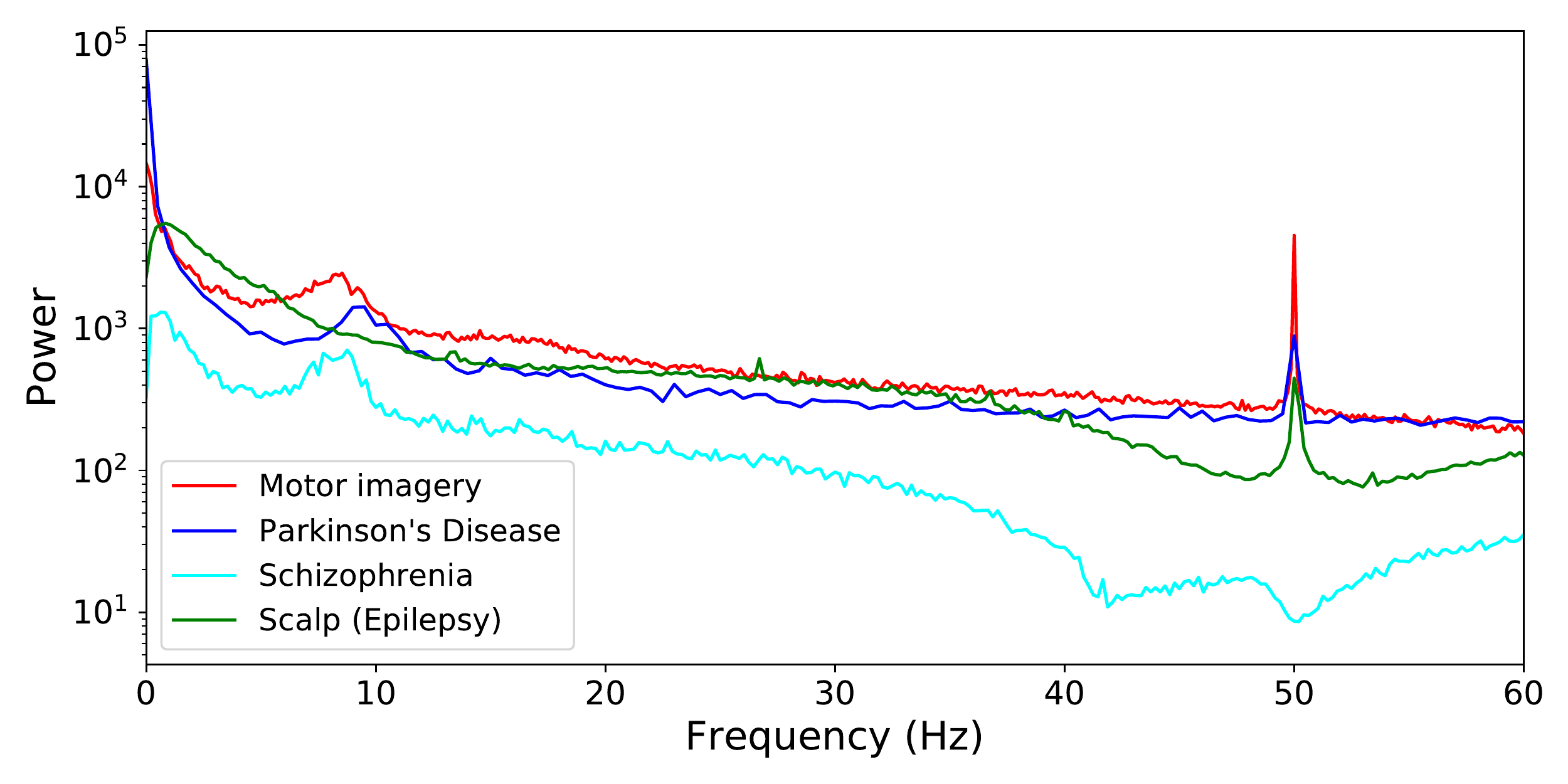}
\caption{Power spectra corresponding to the four considered data sets, averaged over all control subjects. \label{fig:01}}
\end{figure*}

Below are described the four data sets considered in this study; additionally, Tab. \ref{tab:SynthFeatures} reports their main characteristics, and Fig. \ref{fig:01} the corresponding power spectra for control subjects. Unless otherwise specified, no further processing has been performed, i.e. the whole broadband signal has been considered without additional noise reduction or artefact elimination steps.

\begin{table*}[!tb]
\centering
\begin{tabular}{|l|c|c|c|c|c|c|}
\hline
Data set & \# controls & \# patients & Eyes open/close & \# channels & Resolution & Length \\ \hline
Motor Imagery & $110$ & $0$ & Yes / Yes & $64$ & $160$Hz & $1$m \\
Parkinson's disease & $22$ & 74 & Yes / Yes & $32$ & $500$Hz & $3$m \\
Scalp (Epilepsy) & $92$ & $92$ & Yes / No & $22$ & $256$Hz & $>30$s \\
Schizophrenia & $14$ & $14$ & No / Yes & $19$ & $250$Hz & $15$m \\ \hline
\end{tabular}
\caption{Main characteristics of the considered EEG data sets. See Section \ref{sec:datasets} of the main text for details.}
\label{tab:SynthFeatures}
\end{table*}

\subsubsection{Motor Movement/Imagery data set}

This EEG data set is described in \citep{schalk2004bci2000}, and can be downloaded from \url{https://www.physionet.org/pn4/eegmmidb/} \citep{goldberger2000physiobank}. The full data set comprises recordings of subjects performing different motor/imagery tasks, albeit only the eyes open/closed resting states conditions are here considered. A total of $110$ trials (one per subject) are available, recorded with a 64-channel EEG (BCI2000 system). The 64 electrodes were located as per the international 10-10 system, excluding electrodes Nz, F9, F10, FT9, FT10, A1, A2, TP9, TP10, P9, and P10.

\subsubsection{Parkinson's disease data set}
The EEG data set of Parkinson's patients was recorded at Istanbul Medipol University Hospital in Istanbul. PD patients were diagnosed according to the criteria of ``United Kingdom Parkinson's Disease Society Brain Bank'' \citep{daniel1993parkinson}. The Unified Parkinson's Disease Rating Scale (UPDRS) \citep{lang1989assessment} was used in order to determine the clinical features of PD; and the Hoehn-Yahr scale \citep{hoehn1967parkinsonism} was used to determine the disease stage. All patients with PD were evaluated 60 to 90 minutes after their morning dose of levodopa for the EEG recordings. EEG of all healthy controls and Parkinson's Disease patients were recorded in a dimly isolated room. EEG was recorded according to 10-20 system with Brain Amp 32-channel DC system machine from 32 different electrodes. The EEG was recorded with a sampling rate of $500$Hz and with band limits of $0.01-250$Hz. All impedances were kept below $10$kohm and two earlobe electrodes (A1-A2) served as reference electrodes.

\subsubsection{Scalp (epilepsy) data set}

The CHB-MIT Scalp EEG data set is described in \citep{shoeb2009application} and is available for download at \url{https://www.physionet.org/pn6/chbmit/} \citep{goldberger2000physiobank}.
It consists of EEG recordings from paediatric subjects with intractable seizures and free of anti-seizure medication. Note that sub-windows free of seizures are here analysed alongside other groups' control subjects.
All signals were sampled at $256$Hz with $23$ sensors, located according to the International 10-20 system.
Note that Ref. \citep{shoeb2009application} provides no information about the eyes status while recording; in what follows we suppose that all data correspond to an eyes open resting state condition.
As seizures can be of short duration, and for the sake of having time series of similar characteristics across all data sets, only seizure segments longer than $30$ seconds have here been considered, for a total of $92$ instances. The same number of seizure-free segments, of equal duration, have randomly been chosen.

\subsubsection{Schizophrenia data set}

This data set includes resting state EEG recordings for a set of schizophrenia patients and matched control subjects, as described in Ref. \citep{olejarczyk2017graph} and available at \url{http://dx.doi.org/10.18150/repod.0107441}.
The $14$ patients ($7$ males, $27.9 \pm 3.3$ years, and $7$ females, $28.3 \pm 4.1$ years) met International Classification of Diseases ICD-10 criteria for paranoid schizophrenia (category F20.0). The $14$ corresponding healthy controls were $7$ males, age of $26.8 \pm 2.9$ years, and $7$ females, age of $28.7 \pm 3.4$.
Fifteen minutes of EEG data were recorded during an eyes-closed resting state condition. Data were acquired at $250$Hz using the standard 10-20 EEG montage with 19 EEG channels: Fp1, Fp2, F7, F3, Fz, F4, F8, T3, C3, Cz, C4, T4, T5, P3, Pz, P4, T6, O1, O2. The reference electrode was placed at FCz.

\section{Results}

\subsection{Irreversibility of control subjects}

\begin{figure*}[!tb]
\centering
\includegraphics[width=0.9\textwidth]{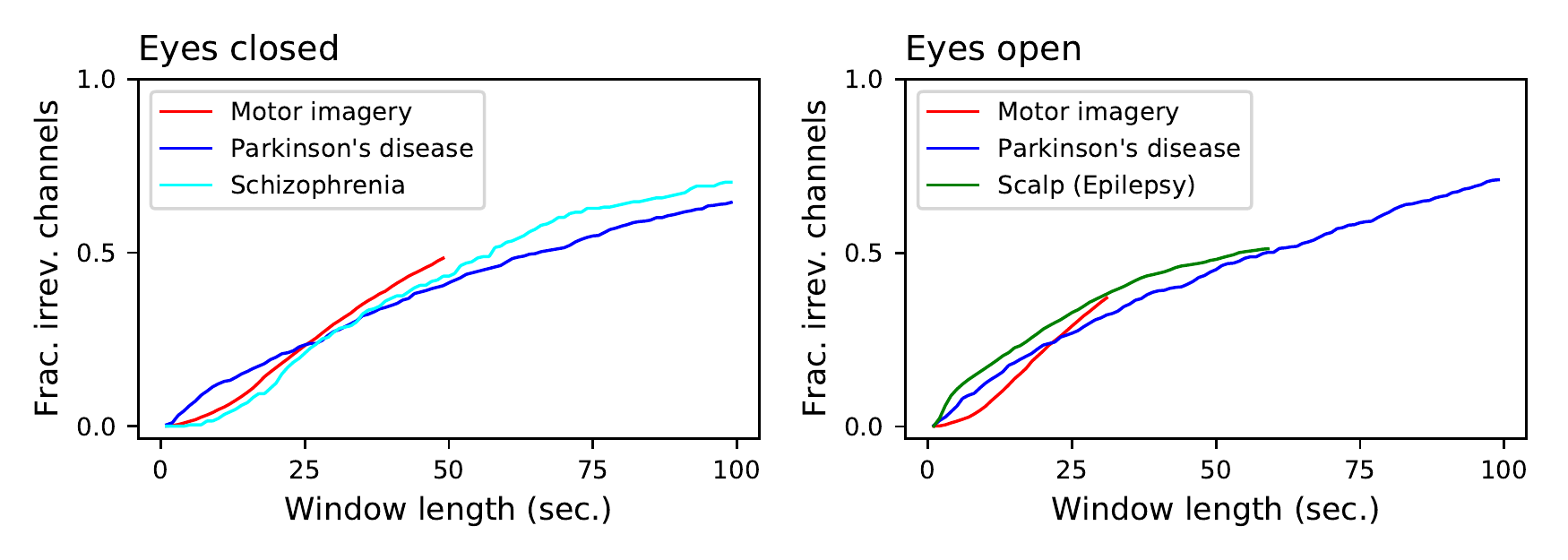}
\caption{Evolution of the fraction of irreversible channels, as a function of the considered window length, for all control subjects. The left and right panels respectively represent data for eyes closed and open resting states. Results here reported correspond to control subjects only, irrespectively of the name in the label - which represents the name of the data set. \label{fig:02}}
\end{figure*}

As a first approach, we calculated how the irreversibility of the healthy (control) brain dynamics evolves as a function of the length of the considered signal. Fig. \ref{fig:02} reports the evolution of the fraction of irreversible time windows, as a function of their length - as described in Section \ref{sec:representing}. Several interesting facts ought to be highlighted.

First of all, all results are quite homogeneous across the considered data sets. This suggests that specific elements, like the used EEG machine, the number of channels or the recording setup have little effect in the metric; and thus that brain irreversibility is a robust property.

\begin{figure*}[!tb]
\centering
\includegraphics[width=0.9\textwidth]{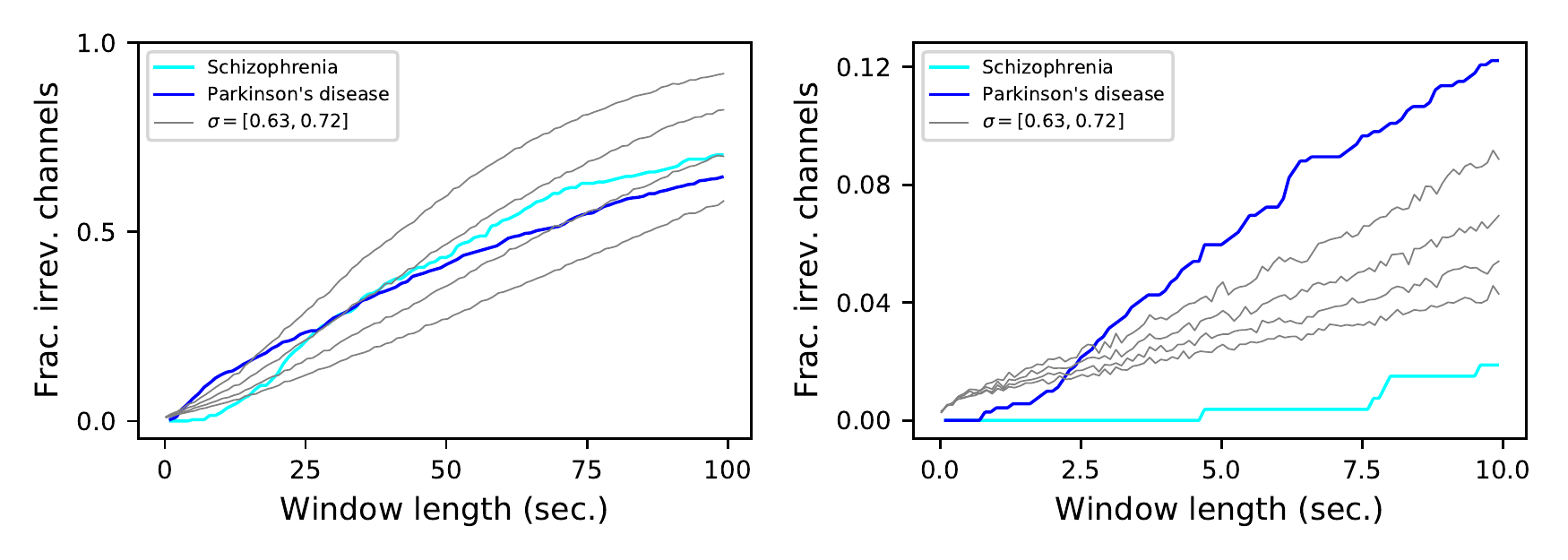}
\caption{Evolution of the fraction of irreversible channels, as a function of the considered window length, for the Schizophrenia and Parkinson's disease data sets, and for the synthetic noisy model (grey lines). From top to bottom, the four grey lines correspond to noise levels of $\sigma = 0.63$, $0.66$, $0.69$, $0.72$. The left and right panels respectively represent the whole results, and a zoom for short window lengths. In all cases, only control subjects have been considered. \label{fig:02Noise}}
\end{figure*}

Secondly, it can be appreciated that the result is a monotonically increasing value with a small slope; even for time windows of $100$ seconds, irreversibility is not detected in about $30\%$ of the cases. The underlying dynamics may thus be irreversible, but a large amount of noise is likely masking such characteristic, so that it can only reliably be detected using long time series. To clarify this point, the left panel of Fig. \ref{fig:02Noise} reports the results for the Parkinson's disease and Schizophrenia data sets (in the eyes closed condition), along with those of the logistic map for different values of additive noise - as defined in Section \ref{sec:noisy}. While the shapes seem {\it prima facie} equal, two important aspects stand out. On one hand, while the irreversibility for the logistic map grows almost linearly with the size of the time window, that of the two EEG data sets seems to grow in a sub-linear way. On the other hand, the behaviour for very short time series is very different, both between the two EEG data sets, and between the EEG data sets and the logistic map - see the magnification in Fig. \ref{fig:02Noise} Right. The observed time series are thus the result of a complex interplay between an irreversible dynamics and observational noise.

\begin{figure*}[!tb]
\centering
\includegraphics[width=0.9\textwidth]{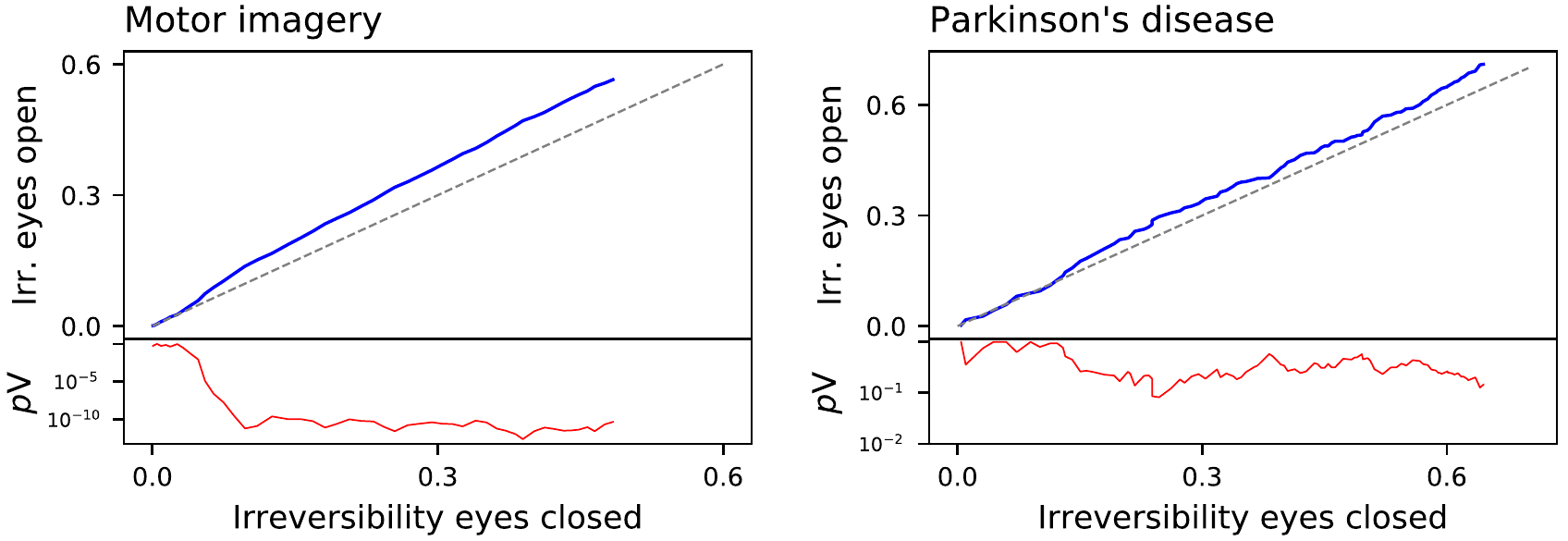}
\caption{Comparison of the fraction of irreversible windows for eyes closed and open conditions, for the motor imaging (left panel) and Parkinson's disease (right panel) data sets. The red lines (right Y axes) depict the evolution of the $\log_{10}$ of the $p$-value of a binomial test, testing if both values are equal - also represented by the grey diagonal line. \label{fig:03}}
\end{figure*}

We then analysed differences in irreversibility between the eyes open and closed conditions. Fig. \ref{fig:03} reports the evolution of the fraction of irreversible windows in the eyes open condition, as a function of the fraction for the eyes closed one.
Each graph is constructed by searching, for a point of coordinates $(x, y)$, the minimum window length for which the fraction of irreversible time series in the eyes closed condition is equal or greater than $x$; then $y$ is set equal to the fraction of irreversible time series in the eyes open condition for that same window length. Points above the main diagonal (dashed grey line) thus indicate that, for a same window length, brain dynamcis is more irreversible in the eyes open condition.

The left and right panels of Fig. \ref{fig:03} respectively report the results corresponding to the motor imaging and Parkinson's disease data sets, {\it i.e.} the two for which both conditions were available.
In both cases the line is above the main diagonal, indicating that the brain is more irreversible in the eyes open condition. This is in agreement with the hypothesis that cognitive activity is associated with irreversibility. Even at rest, leaving the eyes open implies a larger amount of inputs to be processed, and hence a higher activity and irreversibility.

\subsection{Change in the irreversibility due to pathological conditions}

\begin{figure*}[!tb]
\centering
\includegraphics[width=0.9\textwidth]{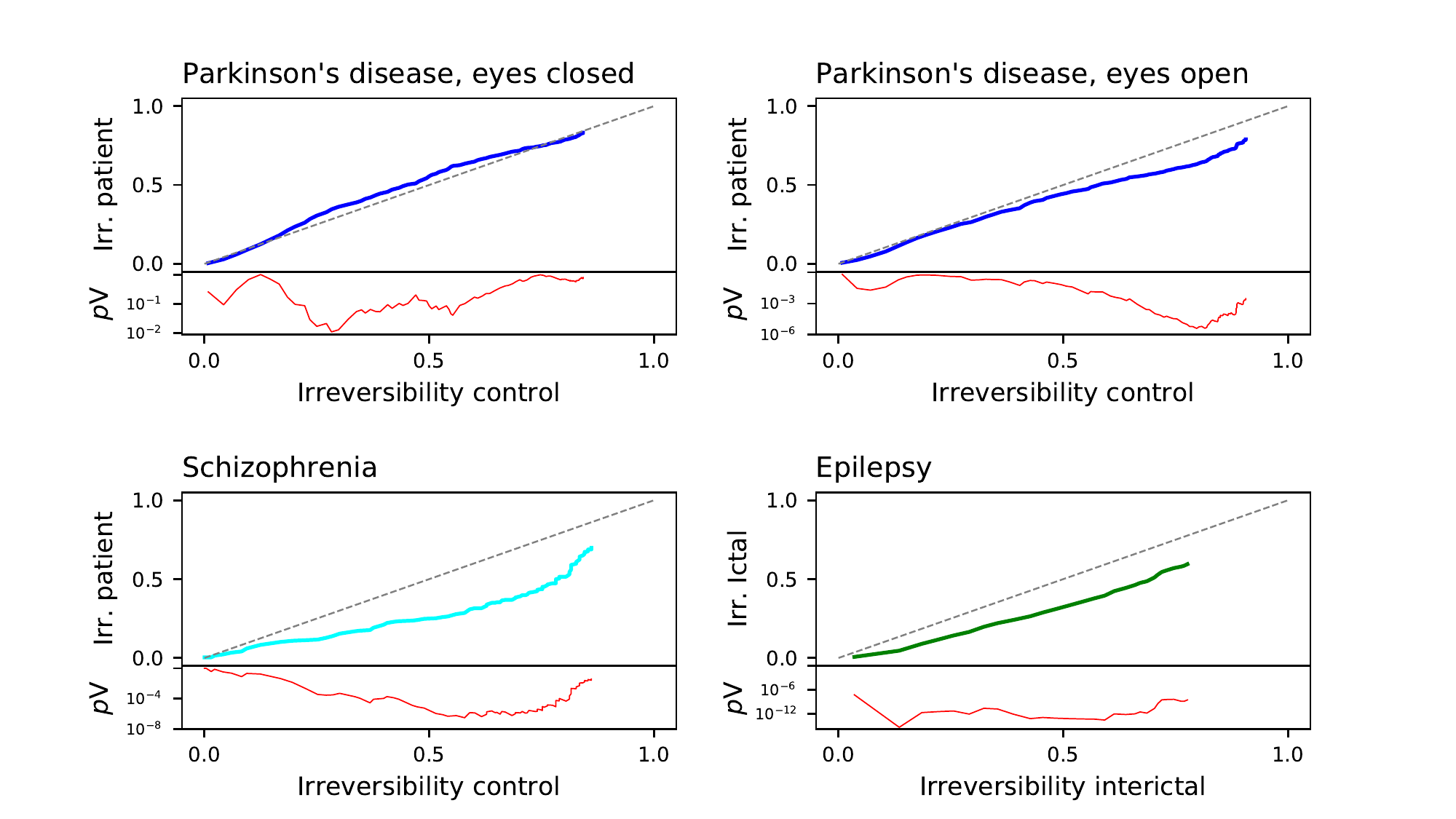}
\caption{Comparison of the fraction of irreversible channels between the patients and the corresponding control subjects of the four considered data sets: Parkinson's disease (with eyes closed and open), Schizophrenia and epilepsy. The red lines depict the evolution of the $\log_{10}$ of the $p$-value of a binomial test, testing if both values are equal - also represented by the grey diagonal line. \label{fig:04}}
\end{figure*}

An interesting question is to understand how different pathologies may affect the irreversibility of the brain, as the latter may yield information about the effect of the former on brain dynamics. Fig. \ref{fig:04} reports the evolution of the irreversibility of patients, as a function of the corresponding irreversibility in the control subjects - note that these graphs have to be interpreted in a way similar to that of Fig. \ref{fig:03}. 

In three of the four data sets, patients exhibit a lower irreversibility, which is especially marked in the case of schizophrenia. These pathologies thus seem to reduce the brain's ability to respond to stimuli; or in other words, that the brain is less prone to deviate from equilibrium. This is nevertheless not homogeneous: while the difference mainly appears for long time series in the Parkinson's disease and the schizophrenia cases, this is not that marked in the case of the epilepsy. This seems to indicate that the brain's dynamical alterations in the two former conditions are identifiable at long time scales, while ictal events are more temporally local.
Parkinson's disease in the eyes closed condition is the exception, displaying a small (non statistically significant) increase in irreversibility. This suggests that, in this pathology, brain dynamics differs in the two conditions, being the irreversibility only different in the eyes open one. This effect may be the result of the visual misperceptions and hallucinations characterising this pathology, which may have a lower impact in eyes closed conditions \citep{davidsdottir2005visual, shine2011visual}.

\begin{figure*}[!tb]
\centering
\includegraphics[width=0.9\textwidth]{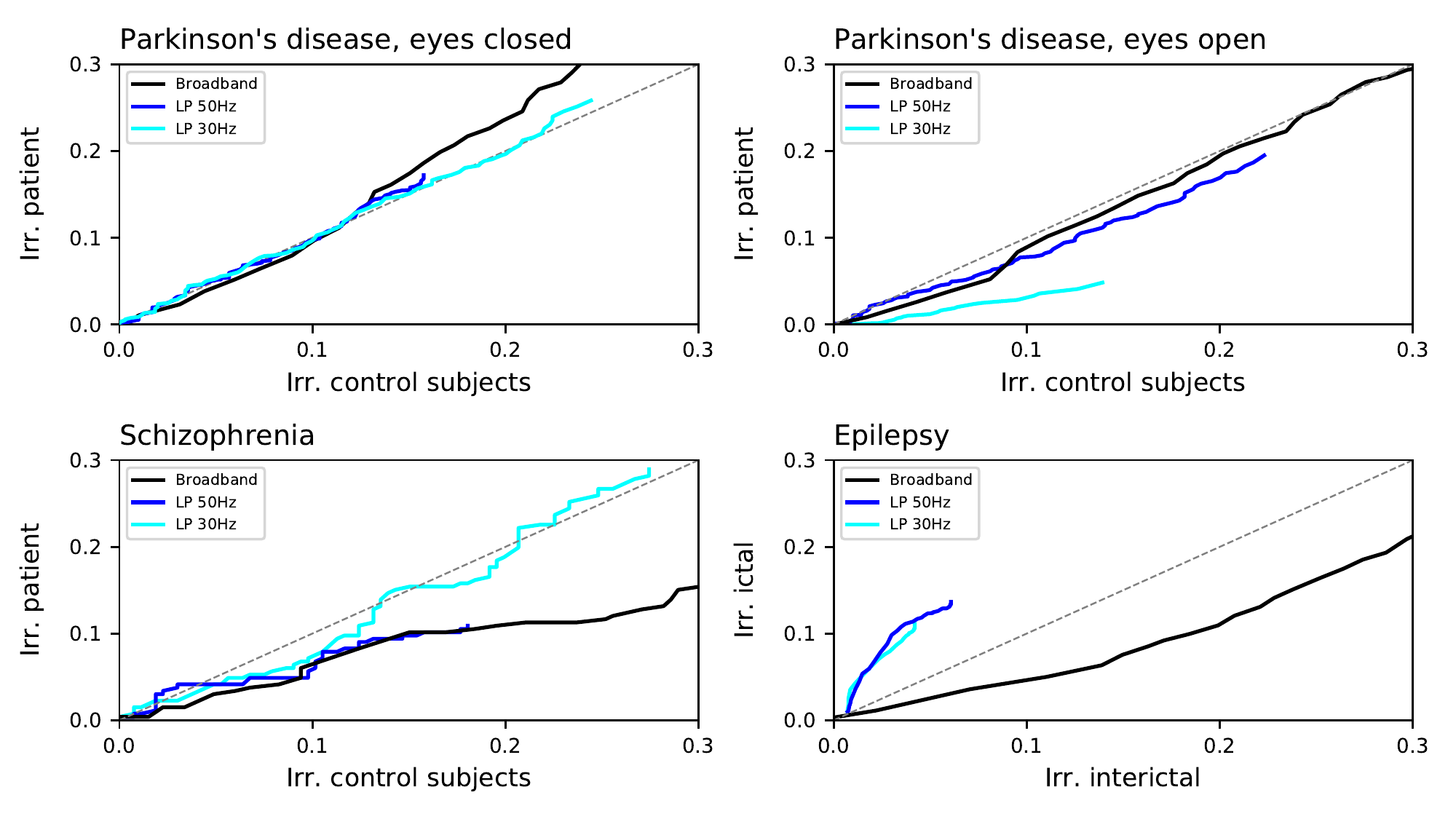}
\caption{Comparison of the fraction of irreversible channels between the patients and the corresponding control subjects of the four considered data sets: Parkinson's disease (with eyes closed and open), Schizophrenia and epilepsy. Black lines correspond to the broadband signals, as reported in Fig. \ref{fig:04}; blue and aqua lines to the signals filtered with respectively a low-pass filter at 50Hz and 30Hz. \label{fig:Filtered}}
\end{figure*}

We further study if these differences between control subjects and patients are consistent across all frequencies, or are specific to some bands. Note that such analysis is also required to exclude that the irreversibility is just a spurious result coming from artefacts or muscular movements. Towards this aim, Fig. \ref{fig:Filtered} depicts three cases: results for the broadband signal (as presented in Fig. \ref{fig:04}), black lines; for signals filtered with a low-pass filter at 50Hz, blue lines; and for signals filtered with a low-pass filter at 30Hz, aqua lines. When the low-pass filter is applied, a corresponding downsampling is also executed, in order to avoid spurious slow dynamics that may bias the irreversibility values.

Results strongly differ for the three data sets. Firstly, in the case of Schizophrenia, applying the filters yields a strong reduction in the difference in irreversibility; on the other hand, the opposite was seen in the case of the Parkinson's disease data set for eyes open, for which the difference between control subjects and patients was substantially increased. Even stronger is the effect of filtering in the case of epilepsy, in which not only the difference between control subjects and patients is increased, but the difference in irreversibility even changed sign.
This suggests the presence of a complex relationship between irreversibility, dynamics at different frequencies and pathologies. In the case of Schizophrenia, patients seem to suffer from reduced irreversibility at high frequencies; while the opposite, i.e. a marked lower reversibility mainly at low frequencies, arises in Parkinson's disease patients.

\begin{figure*}[!tb]
\centering
\includegraphics[width=0.9\textwidth]{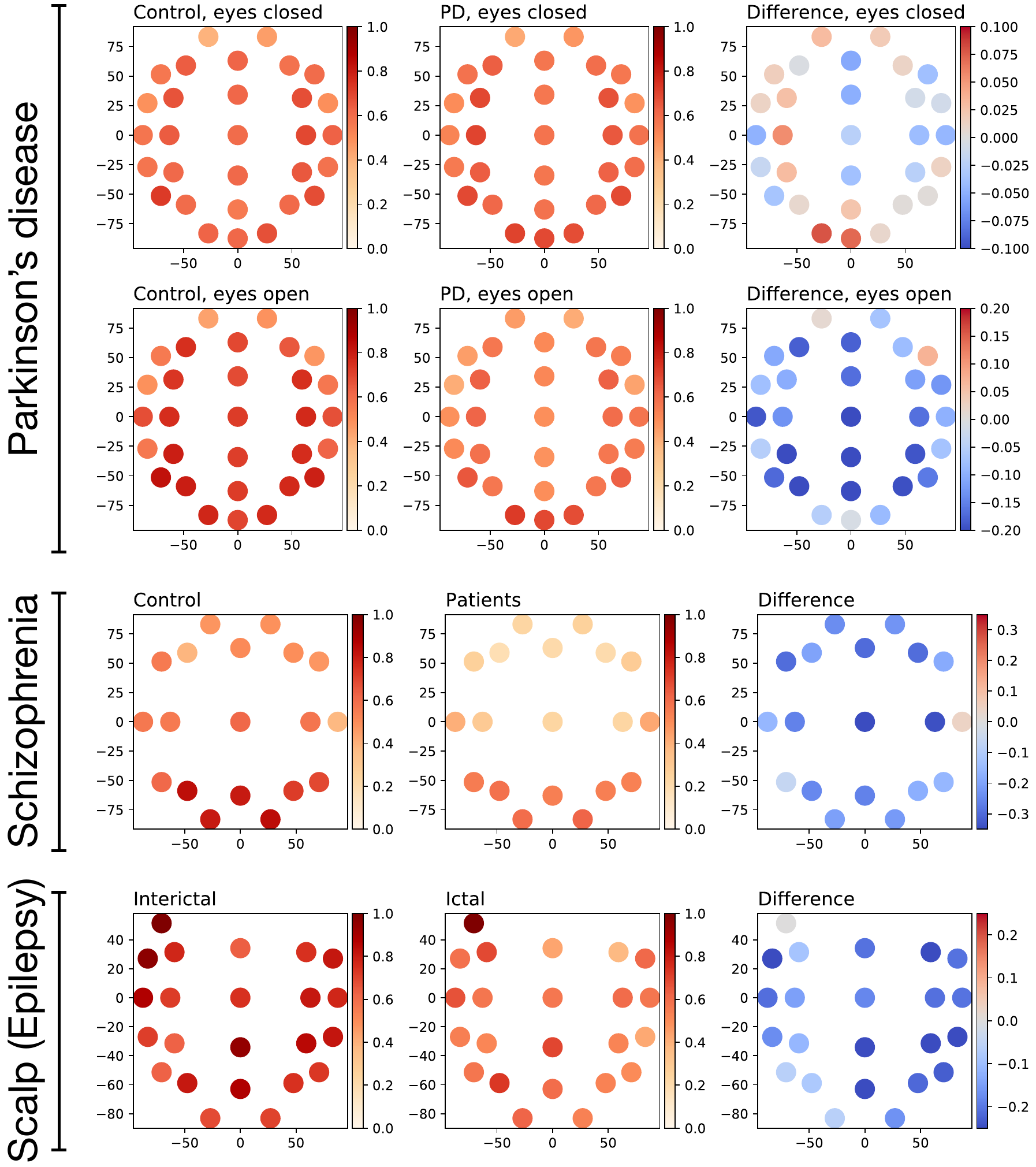}
\caption{Evolution of the average irreversibility by EEG channel in the three data sets corresponding to pathological conditions. Panels in the first and second columns depict the fraction of irreversible windows per channel, the left (right) ones to control subjects (patients). The four right-most panels depict the difference between the patients and the control subjects; positive values (red shapes) indicate a higher irreversibility in patients. \label{fig:05}}
\end{figure*}

We finally analysed how this irreversibility of brain dynamics is spatially distributed throughout the brain in the three pathological conditions here considered. Fig. \ref{fig:05} reports the average irreversibility value according to the EEG sensor, for the broadband signal. This value was calculated by averaging the irreversibility obtained for all window lengths, {\it i.e.} by averaging the curves of Fig. \ref{fig:02}; it therefore represents an overview of the dynamics of the brain at all possible time scales. The four right-most panels of Fig. \ref{fig:05} further report the difference in irreversibility between patients and control subjects - red shades indicating a higher irreversibility in the former. In the case of the Parkinson's disease in eyes closed conditions, patients were characterised by higher irreversibility in the frontal and posterior regions, while this metric was lower in most other regions. In all other cases, the drop in irreversibility characterising patients was more spread, and especially strong on a very extended scalp region, spanning frontal, central and parietal regions.

\subsection{Nature of brain irreversibility}
\label{sec:nature}

As a final issue, we analyse the possible origin of the observed irreversibility. As discussed in Section \ref{sec:irreversibility}, the statistical significance of all presented results has been calculated through the $p$-value of a two-sided binomial test; note that this is equivalent to considering that all values composing the time series are independent, and is thus equivalent to comparing the irreversibility against randomly shuffled series.
We explore another possibility, i.e. the use of IAAFT surrogate time series, which preserve linear autocorrelation and amplitude of the data - see Section \ref{sec:surrogate} for details.
Comparing the results yielded by both approaches allows to partly understand the nature of the observed irreversibility. A statistically significant result in the binomial test suggests the presence of any kind of irreversibility, or of a {\it weak} version of it. If such irreversibility is maintained in the surrogates, it is possibly caused by the linear autocorrelation structure of the time series - as this property is maintained by the IAAFT. On the other hand, if the irreversibility is reduced in the surrogate signals, then the linear autocorrelation can be discarded as a cause - hence indicating a {\it strong} irreversibility.

\begin{figure*}[!tb]
\centering
\includegraphics[width=0.99\textwidth]{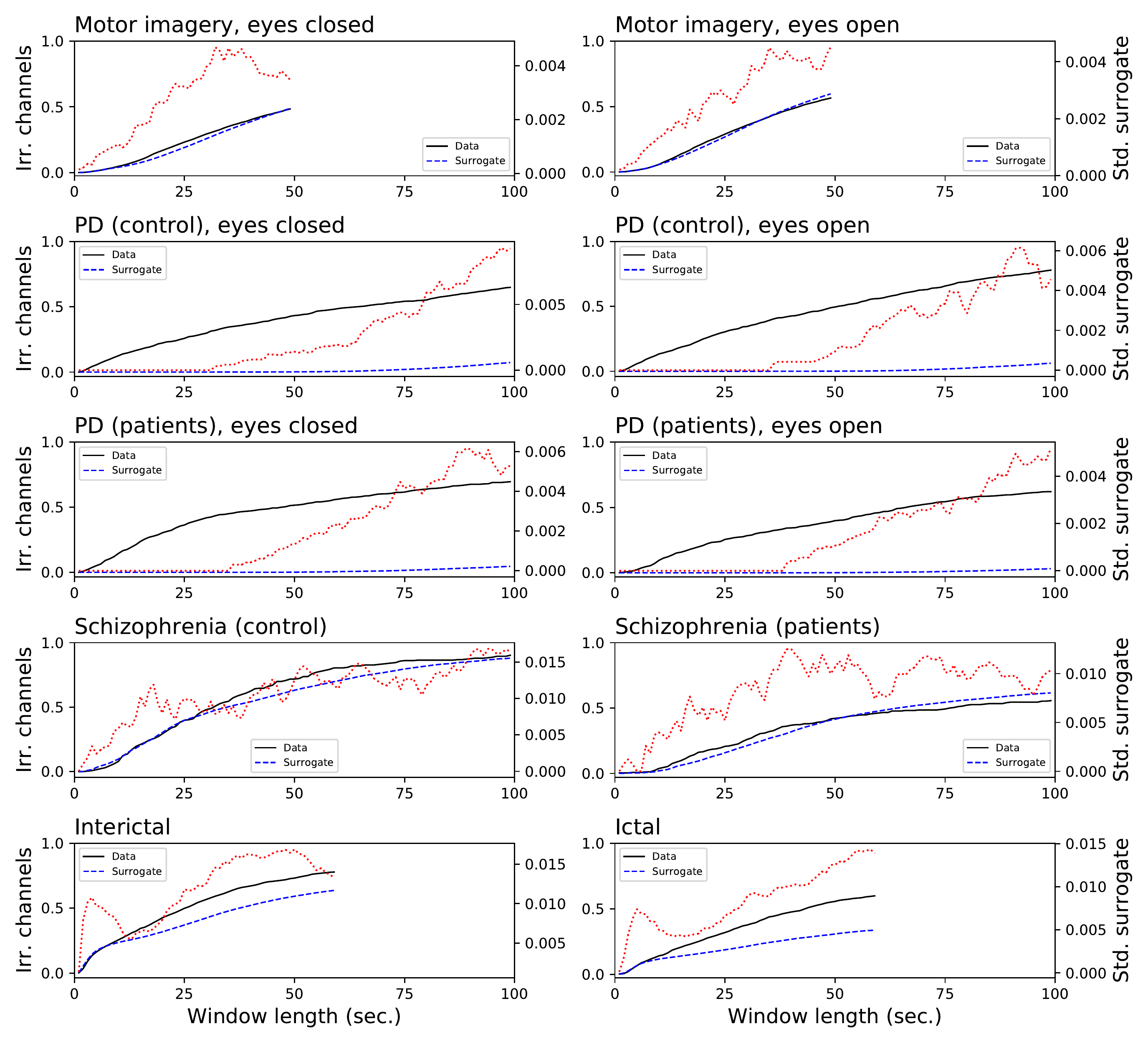}
\caption{Comparison of the fraction of irreversible channels, when the statistical significance is calculated against shuffled time series (black lines) and IAAFT surrogates (blue dashed lines). The red dotted lines (right axes) depict the evolution of the standard deviation of the irreversibility observed in the surrogated time series. \label{fig:08}}
\end{figure*}

Fig. \ref{fig:08} reports the evolution of the irreversibility both in the original time series (black lines) and in the IAAFT surrogates (blue dashed lines). A strong heterogeneity in results can be observed. On one hand, time series in the motor imagery and schizophrenia data sets display a similar irreversibility both in the raw time series and in the surrogate ones, thus indicating that its origin resides in the autocorrelation structure. On the other hand, all cases of the Parkinson's Disease data set can be associated with {\it strong} irreversibility, as this is lost in the surrogates. An intermediate result is finally observed in the case of epilepsy: while inter-ictal windows are more irreversible, ictal ones are characterised by a larger distance from surrogates' irreversibility; this suggests that ictal activity is less irreversible in a {\it weak} sense, but more irreversible in a {\it strong} sense with respect to inter-ictal activity, as already suggested in the literature \citep{van1996time, schindler2016ictal, martinez2018detection}.

\section{Discussion and conclusions}

We used a permutation-entropy based metric to quantify the time-reversal symmetry of spontaneous EEG activity from three groups of patients under two experimental conditions (eyes open and closed). Our results show that resting brain activity is generically time irreversible, and that irreversibility is modulated by simply opening or closing eyes, and altered in a pathology-specific way by psychiatric and neurological disease.

The presence of resting time-reversal asymmetry of electrical activity is consistent with a vision of the brain as a generically out-of-equilibrium system. Our results indicate that at sufficiently long time scales the healthy brain may in fact be operating close to a NESS \citep{livi2013brain}. Moreover, insofar as it has the shape of Eq. \ref{eq:P}'s fluctuation relation, the proposed asymmetry quantifier provides information as to the system's distance from equilibrium. Equilibrium systems fulfil fluctuation-dissipation relations (FDRs). Translated in terms of neural activity, these relations would reflect a substantial equivalence between spontaneous and task-induced brain fluctuations, so that the presence of FDRs would considerably simplify the characterisation of the latter, by allowing to base it merely on the correlation properties of the former \citep{papo2013time}. While the brain as any other biophysical system is not expected to fulfil such equilibrium relations, the extent to which these are violated can nonetheless provide important information on the relation between resting and task-induced activity. The most intuitive way to probe FDR violations would in general consist in comparing correlations of the unperturbed system with stimulus- or generally task-induced ones \citep{martin2001comparison}. However, this method has various shortcoming: 1) it requires separate measures of the correlation and response functions, the latter relying on external perturbations; 2) external perturbations, the effects of which are often difficult to control in a neuroscience context, only represent exogenously promoted cognitive or motor functions; 3) there is no way that perturbations are small enough to guarantee that the measurements are made within the linear response regime. An alternative method to quantify a NESS involves evaluating the property of detailed balance between microstastes of an appropriately coarse-grained mesoscopic representation of the system's dynamics \citep{rupprecht2016fresh}. The proposed method is in some sense a measure of detailed balance violation \citep{zanin2018assessing}, and provides the time-scale-specific magnitude of the distance from equilibrium. 
Finally, while the coarse-graining implicit in both EEG data and in our analyses lose parts of the genuine physical entropy production of the underlying system, the proposed time-irreversibility quantifier can nonetheless be thought to give a lower bound of the system's true one \citep{seifert2019stochastic}.

If time-reversal symmetry reflects a genuine indicator of brain activity efficiency, one would expect that it would vary in a task- and condition-specific manner. Our results show that irreversibility can be modified by an experimental condition as simple as opening and closing eyes (see Figure \ref{fig:03}), consistent with an entropy production interpretation of observed time-reversal symmetry. Our results also generally point to decreased irreversibility in pathology, the lowered proneness to depart from equilibrium being most conspicuous in the schizophrenia group (see Figure \ref{fig:04}). Pathological dynamics seems reminiscent of non-equilibrium systems recovering equilibrium properties at certain scales \citep{egolf2000equilibrium}. In addition, irreversibility patterns showed some degree of pathology-specificity, particularly conspicuous at faster time scales (see Figure \ref{fig:02}). An important general message is then that irreversibility induces a time scale, identified by the transition to irreversibility, both in healthy activity and in pathology. Furthermore, the scales at which irreversibility departs from the healthy pattern also showed some pathology-specificity. 

Time-reversal symmetry alterations showed pathology-specific frequency content (see Figure \ref{fig:05}). This may indicate that the irreversibility pattern consistently seen in healthy controls across different data sets analysed in our study may result from a specific composition of the broad-band frequency spectrum. Conversely, this also suggests that frequency-specific dysfunction associated with various pathologies \citep{lee2001integration, bacsar2008review, roach2013converging, oswal2013synchronized, little2014functional}, usually seen from an exquisitely dynamic view-point, may ultimately affect basic aspects of normal brain efficiency. 

The results of our tests using surrogate time series show a condition-specific dynamical aetiology of time reversibility. In particular, in some pathologies, irreversibility may stem from changes in local linear autocorrelations, while in others it may be a consequence of a different dynamical mechanism, the exact nature of which can only be found by surrogate testing of a different nature from the one used in the present study.
Time irreversibility in the data may be caused by some trivial static nonlinearity rather than by genuine nonlinear dynamics of the system generating the EEG \citep{van1996time}. In our study, the role of additive noise was systematically examined and it was showed to decrease as expected irreversibility (see Figure \ref{fig:02Noise}). On the other hand, when reversibility can be rejected, a static transformation of a linear Gaussian random process can be excluded as an appropriate model for the time series \citep{cox1981statistical}. Evidence abounds for weak non-linearity in multichannel EEG \citep{rombouts1995investigation, pezard1994non} and in the interdepencies between EEG channels \citep{breakspear2002detection, paluvs1996nonlinearity}. However, the role of non-linearity per se in irreversibility may be a complex one, as suggested by increased nonlinearity \citep{pezard2001investigation} but decreased irreversibility in Parkinson's disease in the eyes open (though not in the eyes closed) condition. The frequency-specificity of the irreversibility patterns in the various pathologies considered in the present study may stem from pathology-specific nonlinear features, {\it e.g.} bistability and nondiffusivity, associated with non-Gaussian statistics appearing at certain scales of the underlying dynamics \citep{freyer2009bistability}. Again non-Gaussianity and irreversibility may have a complex, possibly scale-dependent relationship, as the equilibrium systems can exhibit non-Gaussian fluctuations, and conversely non-equilibrium systems can exhibit Gaussian fluctuations.

Finally, the topographically distributed nature of changes in irreversibility with respect to healthy controls would point to diffuse impairment, even for pathologies with localised aetiologies such as Parkinson's disease. Although scalp topographical results should always be interpreted with caution, higher fronto-posterior irreversibility values in Parkinson's disease may point to compensatory mechanisms \citep{blesa2017compensatory}. Altogether, our results may suggest that pathology may change the dynamic process underlying brain dynamics, pushing activity not only towards qualitatively different statistical and dynamical regimes \citep{buiatti2007feedback, papo2014functional}, but also towards different thermodynamical ones.

In conclusion, irreversibility may represent a signature of normal functioning and with the potential to highlight pathology. More generally, the evaluation of irreversibility by comparing the information content of time-reversed processes provides a bridge between dynamics, information and thermodynamics of the brain, and may ultimately help understanding fundamental questions (but otherwise experimentally hard to address) such as information erasure, which is connected to entropy production through Landauer's principles \citep{gaspard2015cycles}.The properties and significance of time scales, the scale- and sampling rate-dependence, aetiology, sensitivity and specificity of time irreversibility will have to be examined with larger and more controlled samples before their clinical significance is corroborated.

\vspace{1.0cm}

\section*{Acknowledgments}
The data collected in Istanbul Medipol University was supported with a T\"{U}B\.{I}TAK Project (No: 214S111).

\section*{Author contributions}
MZ and DP conceived the idea and designed the study. MZ performed all data processing and statistical analyses. BG, TA and LH coordinated the Parkinson's data acquisition. DP and BG interpreted the results. All authors contributed to the final manuscript.

\section*{Competing interests}
No competing interests.

\bibliography{BrainRev.bib}

\begin{thebibliography}{98}%
\makeatletter
\providecommand \@ifxundefined [1]{%
 \@ifx{#1\undefined}
}%
\providecommand \@ifnum [1]{%
 \ifnum #1\expandafter \@firstoftwo
 \else \expandafter \@secondoftwo
 \fi
}%
\providecommand \@ifx [1]{%
 \ifx #1\expandafter \@firstoftwo
 \else \expandafter \@secondoftwo
 \fi
}%
\providecommand \natexlab [1]{#1}%
\providecommand \enquote  [1]{``#1''}%
\providecommand \bibnamefont  [1]{#1}%
\providecommand \bibfnamefont [1]{#1}%
\providecommand \citenamefont [1]{#1}%
\providecommand \href@noop [0]{\@secondoftwo}%
\providecommand \href [0]{\begingroup \@sanitize@url \@href}%
\providecommand \@href[1]{\@@startlink{#1}\@@href}%
\providecommand \@@href[1]{\endgroup#1\@@endlink}%
\providecommand \@sanitize@url [0]{\catcode `\\12\catcode `\$12\catcode
  `\&12\catcode `\#12\catcode `\^12\catcode `\_12\catcode `\%12\relax}%
\providecommand \@@startlink[1]{}%
\providecommand \@@endlink[0]{}%
\providecommand \url  [0]{\begingroup\@sanitize@url \@url }%
\providecommand \@url [1]{\endgroup\@href {#1}{\urlprefix }}%
\providecommand \urlprefix  [0]{URL }%
\providecommand \Eprint [0]{\href }%
\providecommand \doibase [0]{http://dx.doi.org/}%
\providecommand \selectlanguage [0]{\@gobble}%
\providecommand \bibinfo  [0]{\@secondoftwo}%
\providecommand \bibfield  [0]{\@secondoftwo}%
\providecommand \translation [1]{[#1]}%
\providecommand \BibitemOpen [0]{}%
\providecommand \bibitemStop [0]{}%
\providecommand \bibitemNoStop [0]{.\EOS\space}%
\providecommand \EOS [0]{\spacefactor3000\relax}%
\providecommand \BibitemShut  [1]{\csname bibitem#1\endcsname}%
\let\auto@bib@innerbib\@empty
%</preamble>
\bibitem [{\citenamefont {Arieli}\ \emph {et~al.}(1996)\citenamefont {Arieli},
  \citenamefont {Sterkin}, \citenamefont {Grinvald},\ and\ \citenamefont
  {Aertsen}}]{arieli1996dynamics}%
  \BibitemOpen
  \bibfield  {author} {\bibinfo {author} {\bibfnamefont {A.}~\bibnamefont
  {Arieli}}, \bibinfo {author} {\bibfnamefont {A.}~\bibnamefont {Sterkin}},
  \bibinfo {author} {\bibfnamefont {A.}~\bibnamefont {Grinvald}}, \ and\
  \bibinfo {author} {\bibfnamefont {A.}~\bibnamefont {Aertsen}},\ }\href@noop
  {} {\bibfield  {journal} {\bibinfo  {journal} {Science}\ }\textbf {\bibinfo
  {volume} {273}},\ \bibinfo {pages} {1868} (\bibinfo {year}
  {1996})}\BibitemShut {NoStop}%
\bibitem [{\citenamefont {Van~de Ville}\ \emph {et~al.}(2010)\citenamefont
  {Van~de Ville}, \citenamefont {Britz},\ and\ \citenamefont
  {Michel}}]{van2010eeg}%
  \BibitemOpen
  \bibfield  {author} {\bibinfo {author} {\bibfnamefont {D.}~\bibnamefont
  {Van~de Ville}}, \bibinfo {author} {\bibfnamefont {J.}~\bibnamefont {Britz}},
  \ and\ \bibinfo {author} {\bibfnamefont {C.~M.}\ \bibnamefont {Michel}},\
  }\href@noop {} {\bibfield  {journal} {\bibinfo  {journal} {Proceedings of the
  National Academy of Sciences}\ ,\ \bibinfo {pages} {201007841}} (\bibinfo
  {year} {2010})}\BibitemShut {NoStop}%
\bibitem [{\citenamefont {Deco}\ \emph {et~al.}(2011)\citenamefont {Deco},
  \citenamefont {Jirsa},\ and\ \citenamefont {McIntosh}}]{deco2011emerging}%
  \BibitemOpen
  \bibfield  {author} {\bibinfo {author} {\bibfnamefont {G.}~\bibnamefont
  {Deco}}, \bibinfo {author} {\bibfnamefont {V.~K.}\ \bibnamefont {Jirsa}}, \
  and\ \bibinfo {author} {\bibfnamefont {A.~R.}\ \bibnamefont {McIntosh}},\
  }\href@noop {} {\bibfield  {journal} {\bibinfo  {journal} {Nature Reviews
  Neuroscience}\ }\textbf {\bibinfo {volume} {12}},\ \bibinfo {pages} {43}
  (\bibinfo {year} {2011})}\BibitemShut {NoStop}%
\bibitem [{\citenamefont {Kenet}\ \emph {et~al.}(2003)\citenamefont {Kenet},
  \citenamefont {Bibitchkov}, \citenamefont {Tsodyks}, \citenamefont
  {Grinvald},\ and\ \citenamefont {Arieli}}]{kenet2003spontaneously}%
  \BibitemOpen
  \bibfield  {author} {\bibinfo {author} {\bibfnamefont {T.}~\bibnamefont
  {Kenet}}, \bibinfo {author} {\bibfnamefont {D.}~\bibnamefont {Bibitchkov}},
  \bibinfo {author} {\bibfnamefont {M.}~\bibnamefont {Tsodyks}}, \bibinfo
  {author} {\bibfnamefont {A.}~\bibnamefont {Grinvald}}, \ and\ \bibinfo
  {author} {\bibfnamefont {A.}~\bibnamefont {Arieli}},\ }\href@noop {}
  {\bibfield  {journal} {\bibinfo  {journal} {Nature}\ }\textbf {\bibinfo
  {volume} {425}},\ \bibinfo {pages} {954} (\bibinfo {year}
  {2003})}\BibitemShut {NoStop}%
\bibitem [{\citenamefont {Cossart}\ \emph {et~al.}(2003)\citenamefont
  {Cossart}, \citenamefont {Aronov},\ and\ \citenamefont
  {Yuste}}]{cossart2003attractor}%
  \BibitemOpen
  \bibfield  {author} {\bibinfo {author} {\bibfnamefont {R.}~\bibnamefont
  {Cossart}}, \bibinfo {author} {\bibfnamefont {D.}~\bibnamefont {Aronov}}, \
  and\ \bibinfo {author} {\bibfnamefont {R.}~\bibnamefont {Yuste}},\
  }\href@noop {} {\bibfield  {journal} {\bibinfo  {journal} {Nature}\ }\textbf
  {\bibinfo {volume} {423}},\ \bibinfo {pages} {283} (\bibinfo {year}
  {2003})}\BibitemShut {NoStop}%
\bibitem [{\citenamefont {Beggs}\ and\ \citenamefont
  {Plenz}(2004)}]{beggs2004neuronal}%
  \BibitemOpen
  \bibfield  {author} {\bibinfo {author} {\bibfnamefont {J.~M.}\ \bibnamefont
  {Beggs}}\ and\ \bibinfo {author} {\bibfnamefont {D.}~\bibnamefont {Plenz}},\
  }\href@noop {} {\bibfield  {journal} {\bibinfo  {journal} {Journal of
  neuroscience}\ }\textbf {\bibinfo {volume} {24}},\ \bibinfo {pages} {5216}
  (\bibinfo {year} {2004})}\BibitemShut {NoStop}%
\bibitem [{\citenamefont {Ikegaya}\ \emph {et~al.}(2004)\citenamefont
  {Ikegaya}, \citenamefont {Aaron}, \citenamefont {Cossart}, \citenamefont
  {Aronov}, \citenamefont {Lampl}, \citenamefont {Ferster},\ and\ \citenamefont
  {Yuste}}]{ikegaya2004synfire}%
  \BibitemOpen
  \bibfield  {author} {\bibinfo {author} {\bibfnamefont {Y.}~\bibnamefont
  {Ikegaya}}, \bibinfo {author} {\bibfnamefont {G.}~\bibnamefont {Aaron}},
  \bibinfo {author} {\bibfnamefont {R.}~\bibnamefont {Cossart}}, \bibinfo
  {author} {\bibfnamefont {D.}~\bibnamefont {Aronov}}, \bibinfo {author}
  {\bibfnamefont {I.}~\bibnamefont {Lampl}}, \bibinfo {author} {\bibfnamefont
  {D.}~\bibnamefont {Ferster}}, \ and\ \bibinfo {author} {\bibfnamefont
  {R.}~\bibnamefont {Yuste}},\ }\href@noop {} {\bibfield  {journal} {\bibinfo
  {journal} {Science}\ }\textbf {\bibinfo {volume} {304}},\ \bibinfo {pages}
  {559} (\bibinfo {year} {2004})}\BibitemShut {NoStop}%
\bibitem [{\citenamefont {Dragoi}\ and\ \citenamefont
  {Tonegawa}(2011)}]{dragoi2011preplay}%
  \BibitemOpen
  \bibfield  {author} {\bibinfo {author} {\bibfnamefont {G.}~\bibnamefont
  {Dragoi}}\ and\ \bibinfo {author} {\bibfnamefont {S.}~\bibnamefont
  {Tonegawa}},\ }\href@noop {} {\bibfield  {journal} {\bibinfo  {journal}
  {Nature}\ }\textbf {\bibinfo {volume} {469}},\ \bibinfo {pages} {397}
  (\bibinfo {year} {2011})}\BibitemShut {NoStop}%
\bibitem [{\citenamefont {Betzel}\ \emph {et~al.}(2012)\citenamefont {Betzel},
  \citenamefont {Erickson}, \citenamefont {Abell}, \citenamefont {O'Donnell},
  \citenamefont {Hetrick},\ and\ \citenamefont
  {Sporns}}]{betzel2012synchronization}%
  \BibitemOpen
  \bibfield  {author} {\bibinfo {author} {\bibfnamefont {R.~F.}\ \bibnamefont
  {Betzel}}, \bibinfo {author} {\bibfnamefont {M.~A.}\ \bibnamefont
  {Erickson}}, \bibinfo {author} {\bibfnamefont {M.}~\bibnamefont {Abell}},
  \bibinfo {author} {\bibfnamefont {B.~F.}\ \bibnamefont {O'Donnell}}, \bibinfo
  {author} {\bibfnamefont {W.~P.}\ \bibnamefont {Hetrick}}, \ and\ \bibinfo
  {author} {\bibfnamefont {O.}~\bibnamefont {Sporns}},\ }\href@noop {}
  {\bibfield  {journal} {\bibinfo  {journal} {Frontiers in computational
  neuroscience}\ }\textbf {\bibinfo {volume} {6}},\ \bibinfo {pages} {74}
  (\bibinfo {year} {2012})}\BibitemShut {NoStop}%
\bibitem [{\citenamefont {Zhang}\ and\ \citenamefont
  {Raichle}(2010)}]{zhang2010disease}%
  \BibitemOpen
  \bibfield  {author} {\bibinfo {author} {\bibfnamefont {D.}~\bibnamefont
  {Zhang}}\ and\ \bibinfo {author} {\bibfnamefont {M.~E.}\ \bibnamefont
  {Raichle}},\ }\href@noop {} {\bibfield  {journal} {\bibinfo  {journal}
  {Nature Reviews Neurology}\ }\textbf {\bibinfo {volume} {6}},\ \bibinfo
  {pages} {15} (\bibinfo {year} {2010})}\BibitemShut {NoStop}%
\bibitem [{\citenamefont {Alderson-Day}\ \emph {et~al.}(2015)\citenamefont
  {Alderson-Day}, \citenamefont {McCarthy-Jones},\ and\ \citenamefont
  {Fernyhough}}]{alderson2015hearing}%
  \BibitemOpen
  \bibfield  {author} {\bibinfo {author} {\bibfnamefont {B.}~\bibnamefont
  {Alderson-Day}}, \bibinfo {author} {\bibfnamefont {S.}~\bibnamefont
  {McCarthy-Jones}}, \ and\ \bibinfo {author} {\bibfnamefont {C.}~\bibnamefont
  {Fernyhough}},\ }\href@noop {} {\bibfield  {journal} {\bibinfo  {journal}
  {Neuroscience \& Biobehavioral Reviews}\ }\textbf {\bibinfo {volume} {55}},\
  \bibinfo {pages} {78} (\bibinfo {year} {2015})}\BibitemShut {NoStop}%
\bibitem [{\citenamefont {Hohenfeld}\ \emph {et~al.}(2018)\citenamefont
  {Hohenfeld}, \citenamefont {Werner},\ and\ \citenamefont
  {Reetz}}]{hohenfeld2018resting}%
  \BibitemOpen
  \bibfield  {author} {\bibinfo {author} {\bibfnamefont {C.}~\bibnamefont
  {Hohenfeld}}, \bibinfo {author} {\bibfnamefont {C.~J.}\ \bibnamefont
  {Werner}}, \ and\ \bibinfo {author} {\bibfnamefont {K.}~\bibnamefont
  {Reetz}},\ }\href@noop {} {\bibfield  {journal} {\bibinfo  {journal}
  {NeuroImage: Clinical}\ } (\bibinfo {year} {2018})}\BibitemShut {NoStop}%
\bibitem [{\citenamefont {Shew}\ \emph {et~al.}(2009)\citenamefont {Shew},
  \citenamefont {Yang}, \citenamefont {Petermann}, \citenamefont {Roy},\ and\
  \citenamefont {Plenz}}]{shew2009neuronal}%
  \BibitemOpen
  \bibfield  {author} {\bibinfo {author} {\bibfnamefont {W.~L.}\ \bibnamefont
  {Shew}}, \bibinfo {author} {\bibfnamefont {H.}~\bibnamefont {Yang}}, \bibinfo
  {author} {\bibfnamefont {T.}~\bibnamefont {Petermann}}, \bibinfo {author}
  {\bibfnamefont {R.}~\bibnamefont {Roy}}, \ and\ \bibinfo {author}
  {\bibfnamefont {D.}~\bibnamefont {Plenz}},\ }\href@noop {} {\bibfield
  {journal} {\bibinfo  {journal} {Journal of neuroscience}\ }\textbf {\bibinfo
  {volume} {29}},\ \bibinfo {pages} {15595} (\bibinfo {year}
  {2009})}\BibitemShut {NoStop}%
\bibitem [{\citenamefont {Luczak}\ \emph {et~al.}(2009)\citenamefont {Luczak},
  \citenamefont {Barth{\'o}},\ and\ \citenamefont
  {Harris}}]{luczak2009spontaneous}%
  \BibitemOpen
  \bibfield  {author} {\bibinfo {author} {\bibfnamefont {A.}~\bibnamefont
  {Luczak}}, \bibinfo {author} {\bibfnamefont {P.}~\bibnamefont {Barth{\'o}}},
  \ and\ \bibinfo {author} {\bibfnamefont {K.~D.}\ \bibnamefont {Harris}},\
  }\href@noop {} {\bibfield  {journal} {\bibinfo  {journal} {Neuron}\ }\textbf
  {\bibinfo {volume} {62}},\ \bibinfo {pages} {413} (\bibinfo {year}
  {2009})}\BibitemShut {NoStop}%
\bibitem [{\citenamefont {Smith}\ \emph {et~al.}(2009)\citenamefont {Smith},
  \citenamefont {Fox}, \citenamefont {Miller}, \citenamefont {Glahn},
  \citenamefont {Fox}, \citenamefont {Mackay}, \citenamefont {Filippini},
  \citenamefont {Watkins}, \citenamefont {Toro}, \citenamefont {Laird} \emph
  {et~al.}}]{smith2009correspondence}%
  \BibitemOpen
  \bibfield  {author} {\bibinfo {author} {\bibfnamefont {S.~M.}\ \bibnamefont
  {Smith}}, \bibinfo {author} {\bibfnamefont {P.~T.}\ \bibnamefont {Fox}},
  \bibinfo {author} {\bibfnamefont {K.~L.}\ \bibnamefont {Miller}}, \bibinfo
  {author} {\bibfnamefont {D.~C.}\ \bibnamefont {Glahn}}, \bibinfo {author}
  {\bibfnamefont {P.~M.}\ \bibnamefont {Fox}}, \bibinfo {author} {\bibfnamefont
  {C.~E.}\ \bibnamefont {Mackay}}, \bibinfo {author} {\bibfnamefont
  {N.}~\bibnamefont {Filippini}}, \bibinfo {author} {\bibfnamefont {K.~E.}\
  \bibnamefont {Watkins}}, \bibinfo {author} {\bibfnamefont {R.}~\bibnamefont
  {Toro}}, \bibinfo {author} {\bibfnamefont {A.~R.}\ \bibnamefont {Laird}},
  \emph {et~al.},\ }\href@noop {} {\bibfield  {journal} {\bibinfo  {journal}
  {Proceedings of the National Academy of Sciences}\ }\textbf {\bibinfo
  {volume} {106}},\ \bibinfo {pages} {13040} (\bibinfo {year}
  {2009})}\BibitemShut {NoStop}%
\bibitem [{\citenamefont {Papo}(2014)}]{papo2014functional}%
  \BibitemOpen
  \bibfield  {author} {\bibinfo {author} {\bibfnamefont {D.}~\bibnamefont
  {Papo}},\ }\href@noop {} {\bibfield  {journal} {\bibinfo  {journal}
  {Frontiers in systems neuroscience}\ }\textbf {\bibinfo {volume} {8}},\
  \bibinfo {pages} {112} (\bibinfo {year} {2014})}\BibitemShut {NoStop}%
\bibitem [{\citenamefont {Livi}(2013)}]{livi2013brain}%
  \BibitemOpen
  \bibfield  {author} {\bibinfo {author} {\bibfnamefont {R.}~\bibnamefont
  {Livi}},\ }\href@noop {} {\bibfield  {journal} {\bibinfo  {journal} {Chaos,
  Solitons \& Fractals}\ }\textbf {\bibinfo {volume} {55}},\ \bibinfo {pages}
  {60} (\bibinfo {year} {2013})}\BibitemShut {NoStop}%
\bibitem [{\citenamefont {Novikov}\ \emph {et~al.}(1997)\citenamefont
  {Novikov}, \citenamefont {Novikov}, \citenamefont {Shannahoff-Khalsa},
  \citenamefont {Schwartz},\ and\ \citenamefont {Wright}}]{novikov1997scale}%
  \BibitemOpen
  \bibfield  {author} {\bibinfo {author} {\bibfnamefont {E.}~\bibnamefont
  {Novikov}}, \bibinfo {author} {\bibfnamefont {A.}~\bibnamefont {Novikov}},
  \bibinfo {author} {\bibfnamefont {D.}~\bibnamefont {Shannahoff-Khalsa}},
  \bibinfo {author} {\bibfnamefont {B.}~\bibnamefont {Schwartz}}, \ and\
  \bibinfo {author} {\bibfnamefont {J.}~\bibnamefont {Wright}},\ }\href@noop {}
  {\bibfield  {journal} {\bibinfo  {journal} {Physical Review E}\ }\textbf
  {\bibinfo {volume} {56}},\ \bibinfo {pages} {R2387} (\bibinfo {year}
  {1997})}\BibitemShut {NoStop}%
\bibitem [{\citenamefont {Linkenkaer-Hansen}\ \emph {et~al.}(2001)\citenamefont
  {Linkenkaer-Hansen}, \citenamefont {Nikouline}, \citenamefont {Palva},\ and\
  \citenamefont {Ilmoniemi}}]{linkenkaer2001long}%
  \BibitemOpen
  \bibfield  {author} {\bibinfo {author} {\bibfnamefont {K.}~\bibnamefont
  {Linkenkaer-Hansen}}, \bibinfo {author} {\bibfnamefont {V.~V.}\ \bibnamefont
  {Nikouline}}, \bibinfo {author} {\bibfnamefont {J.~M.}\ \bibnamefont
  {Palva}}, \ and\ \bibinfo {author} {\bibfnamefont {R.~J.}\ \bibnamefont
  {Ilmoniemi}},\ }\href@noop {} {\bibfield  {journal} {\bibinfo  {journal}
  {Journal of Neuroscience}\ }\textbf {\bibinfo {volume} {21}},\ \bibinfo
  {pages} {1370} (\bibinfo {year} {2001})}\BibitemShut {NoStop}%
\bibitem [{\citenamefont {Gong}\ \emph {et~al.}(2007)\citenamefont {Gong},
  \citenamefont {Nikolaev},\ and\ \citenamefont {van
  Leeuwen}}]{gong2007intermittent}%
  \BibitemOpen
  \bibfield  {author} {\bibinfo {author} {\bibfnamefont {P.}~\bibnamefont
  {Gong}}, \bibinfo {author} {\bibfnamefont {A.~R.}\ \bibnamefont {Nikolaev}},
  \ and\ \bibinfo {author} {\bibfnamefont {C.}~\bibnamefont {van Leeuwen}},\
  }\href@noop {} {\bibfield  {journal} {\bibinfo  {journal} {Physical Review
  E}\ }\textbf {\bibinfo {volume} {76}},\ \bibinfo {pages} {011904} (\bibinfo
  {year} {2007})}\BibitemShut {NoStop}%
\bibitem [{\citenamefont {Bianco}\ \emph {et~al.}(2007)\citenamefont {Bianco},
  \citenamefont {Ignaccolo}, \citenamefont {Rider}, \citenamefont {Ross},
  \citenamefont {Winsor},\ and\ \citenamefont {Grigolini}}]{bianco2007brain}%
  \BibitemOpen
  \bibfield  {author} {\bibinfo {author} {\bibfnamefont {S.}~\bibnamefont
  {Bianco}}, \bibinfo {author} {\bibfnamefont {M.}~\bibnamefont {Ignaccolo}},
  \bibinfo {author} {\bibfnamefont {M.~S.}\ \bibnamefont {Rider}}, \bibinfo
  {author} {\bibfnamefont {M.~J.}\ \bibnamefont {Ross}}, \bibinfo {author}
  {\bibfnamefont {P.}~\bibnamefont {Winsor}}, \ and\ \bibinfo {author}
  {\bibfnamefont {P.}~\bibnamefont {Grigolini}},\ }\href@noop {} {\bibfield
  {journal} {\bibinfo  {journal} {Physical Review E}\ }\textbf {\bibinfo
  {volume} {75}},\ \bibinfo {pages} {061911} (\bibinfo {year}
  {2007})}\BibitemShut {NoStop}%
\bibitem [{\citenamefont {Freyer}\ \emph {et~al.}(2009)\citenamefont {Freyer},
  \citenamefont {Aquino}, \citenamefont {Robinson}, \citenamefont {Ritter},\
  and\ \citenamefont {Breakspear}}]{freyer2009bistability}%
  \BibitemOpen
  \bibfield  {author} {\bibinfo {author} {\bibfnamefont {F.}~\bibnamefont
  {Freyer}}, \bibinfo {author} {\bibfnamefont {K.}~\bibnamefont {Aquino}},
  \bibinfo {author} {\bibfnamefont {P.~A.}\ \bibnamefont {Robinson}}, \bibinfo
  {author} {\bibfnamefont {P.}~\bibnamefont {Ritter}}, \ and\ \bibinfo {author}
  {\bibfnamefont {M.}~\bibnamefont {Breakspear}},\ }\href@noop {} {\bibfield
  {journal} {\bibinfo  {journal} {Journal of Neuroscience}\ }\textbf {\bibinfo
  {volume} {29}},\ \bibinfo {pages} {8512} (\bibinfo {year}
  {2009})}\BibitemShut {NoStop}%
\bibitem [{\citenamefont {Expert}\ \emph {et~al.}(2011)\citenamefont {Expert},
  \citenamefont {Lambiotte}, \citenamefont {Chialvo}, \citenamefont
  {Christensen}, \citenamefont {Jensen}, \citenamefont {Sharp},\ and\
  \citenamefont {Turkheimer}}]{expert2011self}%
  \BibitemOpen
  \bibfield  {author} {\bibinfo {author} {\bibfnamefont {P.}~\bibnamefont
  {Expert}}, \bibinfo {author} {\bibfnamefont {R.}~\bibnamefont {Lambiotte}},
  \bibinfo {author} {\bibfnamefont {D.~R.}\ \bibnamefont {Chialvo}}, \bibinfo
  {author} {\bibfnamefont {K.}~\bibnamefont {Christensen}}, \bibinfo {author}
  {\bibfnamefont {H.~J.}\ \bibnamefont {Jensen}}, \bibinfo {author}
  {\bibfnamefont {D.~J.}\ \bibnamefont {Sharp}}, \ and\ \bibinfo {author}
  {\bibfnamefont {F.}~\bibnamefont {Turkheimer}},\ }\href@noop {} {\bibfield
  {journal} {\bibinfo  {journal} {Journal of The Royal Society Interface}\
  }\textbf {\bibinfo {volume} {8}},\ \bibinfo {pages} {472} (\bibinfo {year}
  {2011})}\BibitemShut {NoStop}%
\bibitem [{\citenamefont {Wink}\ \emph {et~al.}(2008)\citenamefont {Wink},
  \citenamefont {Bullmore}, \citenamefont {Barnes}, \citenamefont {Bernard},\
  and\ \citenamefont {Suckling}}]{wink2008monofractal}%
  \BibitemOpen
  \bibfield  {author} {\bibinfo {author} {\bibfnamefont {A.-M.}\ \bibnamefont
  {Wink}}, \bibinfo {author} {\bibfnamefont {E.}~\bibnamefont {Bullmore}},
  \bibinfo {author} {\bibfnamefont {A.}~\bibnamefont {Barnes}}, \bibinfo
  {author} {\bibfnamefont {F.}~\bibnamefont {Bernard}}, \ and\ \bibinfo
  {author} {\bibfnamefont {J.}~\bibnamefont {Suckling}},\ }\href@noop {}
  {\bibfield  {journal} {\bibinfo  {journal} {Human brain mapping}\ }\textbf
  {\bibinfo {volume} {29}},\ \bibinfo {pages} {791} (\bibinfo {year}
  {2008})}\BibitemShut {NoStop}%
\bibitem [{\citenamefont {Papo}(2013{\natexlab{a}})}]{papo2013should}%
  \BibitemOpen
  \bibfield  {author} {\bibinfo {author} {\bibfnamefont {D.}~\bibnamefont
  {Papo}},\ }\href@noop {} {\bibfield  {journal} {\bibinfo  {journal}
  {Frontiers in human neuroscience}\ }\textbf {\bibinfo {volume} {7}},\
  \bibinfo {pages} {45} (\bibinfo {year} {2013}{\natexlab{a}})}\BibitemShut
  {NoStop}%
\bibitem [{\citenamefont {Puglisi}\ and\ \citenamefont
  {Villamaina}(2009)}]{puglisi2009irreversible}%
  \BibitemOpen
  \bibfield  {author} {\bibinfo {author} {\bibfnamefont {A.}~\bibnamefont
  {Puglisi}}\ and\ \bibinfo {author} {\bibfnamefont {D.}~\bibnamefont
  {Villamaina}},\ }\href@noop {} {\bibfield  {journal} {\bibinfo  {journal}
  {EPL (Europhysics Letters)}\ }\textbf {\bibinfo {volume} {88}},\ \bibinfo
  {pages} {30004} (\bibinfo {year} {2009})}\BibitemShut {NoStop}%
\bibitem [{\citenamefont {Porporato}\ \emph {et~al.}(2007)\citenamefont
  {Porporato}, \citenamefont {Rigby},\ and\ \citenamefont
  {Daly}}]{porporato2007irreversibility}%
  \BibitemOpen
  \bibfield  {author} {\bibinfo {author} {\bibfnamefont {A.}~\bibnamefont
  {Porporato}}, \bibinfo {author} {\bibfnamefont {J.}~\bibnamefont {Rigby}}, \
  and\ \bibinfo {author} {\bibfnamefont {E.}~\bibnamefont {Daly}},\ }\href@noop
  {} {\bibfield  {journal} {\bibinfo  {journal} {Physical review letters}\
  }\textbf {\bibinfo {volume} {98}},\ \bibinfo {pages} {094101} (\bibinfo
  {year} {2007})}\BibitemShut {NoStop}%
\bibitem [{\citenamefont {Xia}\ \emph {et~al.}(2014)\citenamefont {Xia},
  \citenamefont {Shang}, \citenamefont {Wang},\ and\ \citenamefont
  {Shi}}]{xia2014classifying}%
  \BibitemOpen
  \bibfield  {author} {\bibinfo {author} {\bibfnamefont {J.}~\bibnamefont
  {Xia}}, \bibinfo {author} {\bibfnamefont {P.}~\bibnamefont {Shang}}, \bibinfo
  {author} {\bibfnamefont {J.}~\bibnamefont {Wang}}, \ and\ \bibinfo {author}
  {\bibfnamefont {W.}~\bibnamefont {Shi}},\ }\href@noop {} {\bibfield
  {journal} {\bibinfo  {journal} {Physica A: Statistical Mechanics and Its
  Applications}\ }\textbf {\bibinfo {volume} {400}},\ \bibinfo {pages} {151}
  (\bibinfo {year} {2014})}\BibitemShut {NoStop}%
\bibitem [{\citenamefont {Lawrance}(1991)}]{lawrance1991directionality}%
  \BibitemOpen
  \bibfield  {author} {\bibinfo {author} {\bibfnamefont {A.}~\bibnamefont
  {Lawrance}},\ }\href@noop {} {\bibfield  {journal} {\bibinfo  {journal}
  {International Statistical Review/Revue Internationale de Statistique}\ ,\
  \bibinfo {pages} {67}} (\bibinfo {year} {1991})}\BibitemShut {NoStop}%
\bibitem [{\citenamefont {Weiss}(1975)}]{weiss1975time}%
  \BibitemOpen
  \bibfield  {author} {\bibinfo {author} {\bibfnamefont {G.}~\bibnamefont
  {Weiss}},\ }\href@noop {} {\bibfield  {journal} {\bibinfo  {journal} {Journal
  of Applied Probability}\ }\textbf {\bibinfo {volume} {12}},\ \bibinfo {pages}
  {831} (\bibinfo {year} {1975})}\BibitemShut {NoStop}%
\bibitem [{\citenamefont {Cox}\ \emph {et~al.}(1981)\citenamefont {Cox},
  \citenamefont {Gudmundsson}, \citenamefont {Lindgren}, \citenamefont
  {Bondesson}, \citenamefont {Harsaae}, \citenamefont {Laake}, \citenamefont
  {Juselius},\ and\ \citenamefont {Lauritzen}}]{cox1981statistical}%
  \BibitemOpen
  \bibfield  {author} {\bibinfo {author} {\bibfnamefont {D.~R.}\ \bibnamefont
  {Cox}}, \bibinfo {author} {\bibfnamefont {G.}~\bibnamefont {Gudmundsson}},
  \bibinfo {author} {\bibfnamefont {G.}~\bibnamefont {Lindgren}}, \bibinfo
  {author} {\bibfnamefont {L.}~\bibnamefont {Bondesson}}, \bibinfo {author}
  {\bibfnamefont {E.}~\bibnamefont {Harsaae}}, \bibinfo {author} {\bibfnamefont
  {P.}~\bibnamefont {Laake}}, \bibinfo {author} {\bibfnamefont
  {K.}~\bibnamefont {Juselius}}, \ and\ \bibinfo {author} {\bibfnamefont
  {S.~L.}\ \bibnamefont {Lauritzen}},\ }\href@noop {} {\bibfield  {journal}
  {\bibinfo  {journal} {Scandinavian Journal of Statistics}\ ,\ \bibinfo
  {pages} {93}} (\bibinfo {year} {1981})}\BibitemShut {NoStop}%
\bibitem [{\citenamefont {Stone}\ \emph {et~al.}(1996)\citenamefont {Stone},
  \citenamefont {Landan},\ and\ \citenamefont {May}}]{stone1996detecting}%
  \BibitemOpen
  \bibfield  {author} {\bibinfo {author} {\bibfnamefont {L.}~\bibnamefont
  {Stone}}, \bibinfo {author} {\bibfnamefont {G.}~\bibnamefont {Landan}}, \
  and\ \bibinfo {author} {\bibfnamefont {R.~M.}\ \bibnamefont {May}},\
  }\href@noop {} {\bibfield  {journal} {\bibinfo  {journal} {Proc. R. Soc.
  Lond. B}\ }\textbf {\bibinfo {volume} {263}},\ \bibinfo {pages} {1509}
  (\bibinfo {year} {1996})}\BibitemShut {NoStop}%
\bibitem [{\citenamefont {Gaspard}(2005)}]{gaspard2005brownian}%
  \BibitemOpen
  \bibfield  {author} {\bibinfo {author} {\bibfnamefont {P.}~\bibnamefont
  {Gaspard}},\ }\href@noop {} {\bibfield  {journal} {\bibinfo  {journal} {New
  Journal of Physics}\ }\textbf {\bibinfo {volume} {7}},\ \bibinfo {pages} {77}
  (\bibinfo {year} {2005})}\BibitemShut {NoStop}%
\bibitem [{\citenamefont {Andrieux}\ \emph {et~al.}(2007)\citenamefont
  {Andrieux}, \citenamefont {Gaspard}, \citenamefont {Ciliberto}, \citenamefont
  {Garnier}, \citenamefont {Joubaud},\ and\ \citenamefont
  {Petrosyan}}]{andrieux2007entropy}%
  \BibitemOpen
  \bibfield  {author} {\bibinfo {author} {\bibfnamefont {D.}~\bibnamefont
  {Andrieux}}, \bibinfo {author} {\bibfnamefont {P.}~\bibnamefont {Gaspard}},
  \bibinfo {author} {\bibfnamefont {S.}~\bibnamefont {Ciliberto}}, \bibinfo
  {author} {\bibfnamefont {N.}~\bibnamefont {Garnier}}, \bibinfo {author}
  {\bibfnamefont {S.}~\bibnamefont {Joubaud}}, \ and\ \bibinfo {author}
  {\bibfnamefont {A.}~\bibnamefont {Petrosyan}},\ }\href@noop {} {\bibfield
  {journal} {\bibinfo  {journal} {Physical review letters}\ }\textbf {\bibinfo
  {volume} {98}},\ \bibinfo {pages} {150601} (\bibinfo {year}
  {2007})}\BibitemShut {NoStop}%
\bibitem [{\citenamefont {Rold{\'a}n}\ and\ \citenamefont
  {Parrondo}(2010)}]{roldan2010estimating}%
  \BibitemOpen
  \bibfield  {author} {\bibinfo {author} {\bibfnamefont {{\'E}.}~\bibnamefont
  {Rold{\'a}n}}\ and\ \bibinfo {author} {\bibfnamefont {J.~M.}\ \bibnamefont
  {Parrondo}},\ }\href@noop {} {\bibfield  {journal} {\bibinfo  {journal}
  {Physical review letters}\ }\textbf {\bibinfo {volume} {105}},\ \bibinfo
  {pages} {150607} (\bibinfo {year} {2010})}\BibitemShut {NoStop}%
\bibitem [{\citenamefont {Seifert}(2019)}]{seifert2019stochastic}%
  \BibitemOpen
  \bibfield  {author} {\bibinfo {author} {\bibfnamefont {U.}~\bibnamefont
  {Seifert}},\ }\href@noop {} {\bibfield  {journal} {\bibinfo  {journal}
  {Annual Review of Condensed Matter Physics}\ }\textbf {\bibinfo {volume}
  {10}},\ \bibinfo {pages} {171} (\bibinfo {year} {2019})}\BibitemShut
  {NoStop}%
\bibitem [{\citenamefont {Gaspard}(2004)}]{gaspard2004time}%
  \BibitemOpen
  \bibfield  {author} {\bibinfo {author} {\bibfnamefont {P.}~\bibnamefont
  {Gaspard}},\ }\href@noop {} {\bibfield  {journal} {\bibinfo  {journal}
  {Journal of statistical physics}\ }\textbf {\bibinfo {volume} {117}},\
  \bibinfo {pages} {599} (\bibinfo {year} {2004})}\BibitemShut {NoStop}%
\bibitem [{\citenamefont {Evans}\ \emph {et~al.}(1993)\citenamefont {Evans},
  \citenamefont {Cohen},\ and\ \citenamefont {Morriss}}]{evans1993probability}%
  \BibitemOpen
  \bibfield  {author} {\bibinfo {author} {\bibfnamefont {D.~J.}\ \bibnamefont
  {Evans}}, \bibinfo {author} {\bibfnamefont {E.~G.~D.}\ \bibnamefont {Cohen}},
  \ and\ \bibinfo {author} {\bibfnamefont {G.~P.}\ \bibnamefont {Morriss}},\
  }\href@noop {} {\bibfield  {journal} {\bibinfo  {journal} {Physical review
  letters}\ }\textbf {\bibinfo {volume} {71}},\ \bibinfo {pages} {2401}
  (\bibinfo {year} {1993})}\BibitemShut {NoStop}%
\bibitem [{\citenamefont {Gallavotti}\ and\ \citenamefont
  {Cohen}(1995)}]{gallavotti1995dynamical}%
  \BibitemOpen
  \bibfield  {author} {\bibinfo {author} {\bibfnamefont {G.}~\bibnamefont
  {Gallavotti}}\ and\ \bibinfo {author} {\bibfnamefont {E.~G.~D.}\ \bibnamefont
  {Cohen}},\ }\href@noop {} {\bibfield  {journal} {\bibinfo  {journal} {Journal
  of Statistical Physics}\ }\textbf {\bibinfo {volume} {80}},\ \bibinfo {pages}
  {931} (\bibinfo {year} {1995})}\BibitemShut {NoStop}%
\bibitem [{\citenamefont {Crooks}(2000)}]{crooks2000path}%
  \BibitemOpen
  \bibfield  {author} {\bibinfo {author} {\bibfnamefont {G.~E.}\ \bibnamefont
  {Crooks}},\ }\href@noop {} {\bibfield  {journal} {\bibinfo  {journal}
  {Physical review E}\ }\textbf {\bibinfo {volume} {61}},\ \bibinfo {pages}
  {2361} (\bibinfo {year} {2000})}\BibitemShut {NoStop}%
\bibitem [{\citenamefont {Evans}\ and\ \citenamefont
  {Searles}(2002)}]{evans2002fluctuation}%
  \BibitemOpen
  \bibfield  {author} {\bibinfo {author} {\bibfnamefont {D.~J.}\ \bibnamefont
  {Evans}}\ and\ \bibinfo {author} {\bibfnamefont {D.~J.}\ \bibnamefont
  {Searles}},\ }\href@noop {} {\bibfield  {journal} {\bibinfo  {journal}
  {Advances in Physics}\ }\textbf {\bibinfo {volume} {51}},\ \bibinfo {pages}
  {1529} (\bibinfo {year} {2002})}\BibitemShut {NoStop}%
\bibitem [{\citenamefont {Ramsey}\ and\ \citenamefont
  {Rothman}(1996)}]{ramsey1996time}%
  \BibitemOpen
  \bibfield  {author} {\bibinfo {author} {\bibfnamefont {J.~B.}\ \bibnamefont
  {Ramsey}}\ and\ \bibinfo {author} {\bibfnamefont {P.}~\bibnamefont
  {Rothman}},\ }\href@noop {} {\bibfield  {journal} {\bibinfo  {journal}
  {Journal of Money, Credit and Banking}\ }\textbf {\bibinfo {volume} {28}},\
  \bibinfo {pages} {1} (\bibinfo {year} {1996})}\BibitemShut {NoStop}%
\bibitem [{\citenamefont {Zumbach}(2009)}]{zumbach2009time}%
  \BibitemOpen
  \bibfield  {author} {\bibinfo {author} {\bibfnamefont {G.}~\bibnamefont
  {Zumbach}},\ }\href@noop {} {\bibfield  {journal} {\bibinfo  {journal}
  {Quantitative Finance}\ }\textbf {\bibinfo {volume} {9}},\ \bibinfo {pages}
  {505} (\bibinfo {year} {2009})}\BibitemShut {NoStop}%
\bibitem [{\citenamefont {Costa}\ \emph {et~al.}(2005)\citenamefont {Costa},
  \citenamefont {Goldberger},\ and\ \citenamefont {Peng}}]{costa2005broken}%
  \BibitemOpen
  \bibfield  {author} {\bibinfo {author} {\bibfnamefont {M.}~\bibnamefont
  {Costa}}, \bibinfo {author} {\bibfnamefont {A.~L.}\ \bibnamefont
  {Goldberger}}, \ and\ \bibinfo {author} {\bibfnamefont {C.-K.}\ \bibnamefont
  {Peng}},\ }\href@noop {} {\bibfield  {journal} {\bibinfo  {journal} {Physical
  review letters}\ }\textbf {\bibinfo {volume} {95}},\ \bibinfo {pages}
  {198102} (\bibinfo {year} {2005})}\BibitemShut {NoStop}%
\bibitem [{\citenamefont {Guzik}\ \emph {et~al.}(2006)\citenamefont {Guzik},
  \citenamefont {Piskorski}, \citenamefont {Krauze}, \citenamefont
  {Wykretowicz},\ and\ \citenamefont {Wysocki}}]{guzik2006heart}%
  \BibitemOpen
  \bibfield  {author} {\bibinfo {author} {\bibfnamefont {P.}~\bibnamefont
  {Guzik}}, \bibinfo {author} {\bibfnamefont {J.}~\bibnamefont {Piskorski}},
  \bibinfo {author} {\bibfnamefont {T.}~\bibnamefont {Krauze}}, \bibinfo
  {author} {\bibfnamefont {A.}~\bibnamefont {Wykretowicz}}, \ and\ \bibinfo
  {author} {\bibfnamefont {H.}~\bibnamefont {Wysocki}},\ }\href@noop {}
  {\bibfield  {journal} {\bibinfo  {journal} {Biomedizinische Technik}\
  }\textbf {\bibinfo {volume} {51}},\ \bibinfo {pages} {272} (\bibinfo {year}
  {2006})}\BibitemShut {NoStop}%
\bibitem [{\citenamefont {Piskorski}\ and\ \citenamefont
  {Guzik}(2007)}]{piskorski2007geometry}%
  \BibitemOpen
  \bibfield  {author} {\bibinfo {author} {\bibfnamefont {J.}~\bibnamefont
  {Piskorski}}\ and\ \bibinfo {author} {\bibfnamefont {P.}~\bibnamefont
  {Guzik}},\ }\href@noop {} {\bibfield  {journal} {\bibinfo  {journal}
  {Physiological measurement}\ }\textbf {\bibinfo {volume} {28}},\ \bibinfo
  {pages} {287} (\bibinfo {year} {2007})}\BibitemShut {NoStop}%
\bibitem [{\citenamefont {Porta}\ \emph {et~al.}(2006)\citenamefont {Porta},
  \citenamefont {Guzzetti}, \citenamefont {Montano}, \citenamefont
  {Gnecchi-Ruscone}, \citenamefont {Furlan},\ and\ \citenamefont
  {Malliani}}]{porta2006time}%
  \BibitemOpen
  \bibfield  {author} {\bibinfo {author} {\bibfnamefont {A.}~\bibnamefont
  {Porta}}, \bibinfo {author} {\bibfnamefont {S.}~\bibnamefont {Guzzetti}},
  \bibinfo {author} {\bibfnamefont {N.}~\bibnamefont {Montano}}, \bibinfo
  {author} {\bibfnamefont {T.}~\bibnamefont {Gnecchi-Ruscone}}, \bibinfo
  {author} {\bibfnamefont {R.}~\bibnamefont {Furlan}}, \ and\ \bibinfo {author}
  {\bibfnamefont {A.}~\bibnamefont {Malliani}},\ }in\ \href@noop {} {\emph
  {\bibinfo {booktitle} {2006 Computers in Cardiology}}}\ (\bibinfo
  {organization} {IEEE},\ \bibinfo {year} {2006})\ pp.\ \bibinfo {pages}
  {77--80}\BibitemShut {NoStop}%
\bibitem [{\citenamefont {Porta}\ \emph {et~al.}(2008)\citenamefont {Porta},
  \citenamefont {Casali}, \citenamefont {Casali}, \citenamefont
  {Gnecchi-Ruscone}, \citenamefont {Tobaldini}, \citenamefont {Montano},
  \citenamefont {Lange}, \citenamefont {Geue}, \citenamefont {Cysarz},\ and\
  \citenamefont {Van~Leeuwen}}]{porta2008temporal}%
  \BibitemOpen
  \bibfield  {author} {\bibinfo {author} {\bibfnamefont {A.}~\bibnamefont
  {Porta}}, \bibinfo {author} {\bibfnamefont {K.~R.}\ \bibnamefont {Casali}},
  \bibinfo {author} {\bibfnamefont {A.~G.}\ \bibnamefont {Casali}}, \bibinfo
  {author} {\bibfnamefont {T.}~\bibnamefont {Gnecchi-Ruscone}}, \bibinfo
  {author} {\bibfnamefont {E.}~\bibnamefont {Tobaldini}}, \bibinfo {author}
  {\bibfnamefont {N.}~\bibnamefont {Montano}}, \bibinfo {author} {\bibfnamefont
  {S.}~\bibnamefont {Lange}}, \bibinfo {author} {\bibfnamefont
  {D.}~\bibnamefont {Geue}}, \bibinfo {author} {\bibfnamefont {D.}~\bibnamefont
  {Cysarz}}, \ and\ \bibinfo {author} {\bibfnamefont {P.}~\bibnamefont
  {Van~Leeuwen}},\ }\href@noop {} {\bibfield  {journal} {\bibinfo  {journal}
  {American Journal of Physiology-Regulatory, Integrative and Comparative
  Physiology}\ } (\bibinfo {year} {2008})}\BibitemShut {NoStop}%
\bibitem [{\citenamefont {Porta}\ \emph {et~al.}(2009)\citenamefont {Porta},
  \citenamefont {D'addio}, \citenamefont {Bassani}, \citenamefont {Maestri},\
  and\ \citenamefont {Pinna}}]{porta2009assessment}%
  \BibitemOpen
  \bibfield  {author} {\bibinfo {author} {\bibfnamefont {A.}~\bibnamefont
  {Porta}}, \bibinfo {author} {\bibfnamefont {G.}~\bibnamefont {D'addio}},
  \bibinfo {author} {\bibfnamefont {T.}~\bibnamefont {Bassani}}, \bibinfo
  {author} {\bibfnamefont {R.}~\bibnamefont {Maestri}}, \ and\ \bibinfo
  {author} {\bibfnamefont {G.~D.}\ \bibnamefont {Pinna}},\ }\href@noop {}
  {\bibfield  {journal} {\bibinfo  {journal} {Philosophical Transactions of the
  Royal Society A: Mathematical, Physical and Engineering Sciences}\ }\textbf
  {\bibinfo {volume} {367}},\ \bibinfo {pages} {1359} (\bibinfo {year}
  {2009})}\BibitemShut {NoStop}%
\bibitem [{\citenamefont {Karmakar}\ \emph {et~al.}(2009)\citenamefont
  {Karmakar}, \citenamefont {Khandoker}, \citenamefont {Gubbi},\ and\
  \citenamefont {Palaniswami}}]{karmakar2009defining}%
  \BibitemOpen
  \bibfield  {author} {\bibinfo {author} {\bibfnamefont {C.~K.}\ \bibnamefont
  {Karmakar}}, \bibinfo {author} {\bibfnamefont {A.}~\bibnamefont {Khandoker}},
  \bibinfo {author} {\bibfnamefont {J.}~\bibnamefont {Gubbi}}, \ and\ \bibinfo
  {author} {\bibfnamefont {M.}~\bibnamefont {Palaniswami}},\ }\href@noop {}
  {\bibfield  {journal} {\bibinfo  {journal} {Physiological measurement}\
  }\textbf {\bibinfo {volume} {30}},\ \bibinfo {pages} {1227} (\bibinfo {year}
  {2009})}\BibitemShut {NoStop}%
\bibitem [{\citenamefont {Hou}\ \emph {et~al.}(2010)\citenamefont {Hou},
  \citenamefont {Zhuang}, \citenamefont {Bian}, \citenamefont {Tong},
  \citenamefont {Chen}, \citenamefont {Yin}, \citenamefont {Qiu},\ and\
  \citenamefont {Ning}}]{hou2010analysis}%
  \BibitemOpen
  \bibfield  {author} {\bibinfo {author} {\bibfnamefont {F.}~\bibnamefont
  {Hou}}, \bibinfo {author} {\bibfnamefont {J.}~\bibnamefont {Zhuang}},
  \bibinfo {author} {\bibfnamefont {C.}~\bibnamefont {Bian}}, \bibinfo {author}
  {\bibfnamefont {T.}~\bibnamefont {Tong}}, \bibinfo {author} {\bibfnamefont
  {Y.}~\bibnamefont {Chen}}, \bibinfo {author} {\bibfnamefont {J.}~\bibnamefont
  {Yin}}, \bibinfo {author} {\bibfnamefont {X.}~\bibnamefont {Qiu}}, \ and\
  \bibinfo {author} {\bibfnamefont {X.}~\bibnamefont {Ning}},\ }\href@noop {}
  {\bibfield  {journal} {\bibinfo  {journal} {Physica A: Statistical Mechanics
  and its Applications}\ }\textbf {\bibinfo {volume} {389}},\ \bibinfo {pages}
  {754} (\bibinfo {year} {2010})}\BibitemShut {NoStop}%
\bibitem [{\citenamefont {Timmer}\ \emph {et~al.}(1993)\citenamefont {Timmer},
  \citenamefont {Gantert}, \citenamefont {Deuschl},\ and\ \citenamefont
  {Honerkamp}}]{timmer1993characteristics}%
  \BibitemOpen
  \bibfield  {author} {\bibinfo {author} {\bibfnamefont {J.}~\bibnamefont
  {Timmer}}, \bibinfo {author} {\bibfnamefont {C.}~\bibnamefont {Gantert}},
  \bibinfo {author} {\bibfnamefont {G.}~\bibnamefont {Deuschl}}, \ and\
  \bibinfo {author} {\bibfnamefont {J.}~\bibnamefont {Honerkamp}},\ }\href@noop
  {} {\bibfield  {journal} {\bibinfo  {journal} {Biological cybernetics}\
  }\textbf {\bibinfo {volume} {70}},\ \bibinfo {pages} {75} (\bibinfo {year}
  {1993})}\BibitemShut {NoStop}%
\bibitem [{\citenamefont {Palu{\v{s}}}(1996)}]{paluvs1996nonlinearity}%
  \BibitemOpen
  \bibfield  {author} {\bibinfo {author} {\bibfnamefont {M.}~\bibnamefont
  {Palu{\v{s}}}},\ }\href@noop {} {\bibfield  {journal} {\bibinfo  {journal}
  {Biological cybernetics}\ }\textbf {\bibinfo {volume} {75}},\ \bibinfo
  {pages} {389} (\bibinfo {year} {1996})}\BibitemShut {NoStop}%
\bibitem [{\citenamefont {Van~der Heyden}\ \emph {et~al.}(1996)\citenamefont
  {Van~der Heyden}, \citenamefont {Diks}, \citenamefont {Pijn},\ and\
  \citenamefont {Velis}}]{van1996time}%
  \BibitemOpen
  \bibfield  {author} {\bibinfo {author} {\bibfnamefont {M.}~\bibnamefont
  {Van~der Heyden}}, \bibinfo {author} {\bibfnamefont {C.}~\bibnamefont
  {Diks}}, \bibinfo {author} {\bibfnamefont {J.}~\bibnamefont {Pijn}}, \ and\
  \bibinfo {author} {\bibfnamefont {D.}~\bibnamefont {Velis}},\ }\href@noop {}
  {\bibfield  {journal} {\bibinfo  {journal} {Physics Letters A}\ }\textbf
  {\bibinfo {volume} {216}},\ \bibinfo {pages} {283} (\bibinfo {year}
  {1996})}\BibitemShut {NoStop}%
\bibitem [{\citenamefont {Ehlers}\ \emph {et~al.}(1998)\citenamefont {Ehlers},
  \citenamefont {Havstad}, \citenamefont {Prichard},\ and\ \citenamefont
  {Theiler}}]{ehlers1998low}%
  \BibitemOpen
  \bibfield  {author} {\bibinfo {author} {\bibfnamefont {C.~L.}\ \bibnamefont
  {Ehlers}}, \bibinfo {author} {\bibfnamefont {J.}~\bibnamefont {Havstad}},
  \bibinfo {author} {\bibfnamefont {D.}~\bibnamefont {Prichard}}, \ and\
  \bibinfo {author} {\bibfnamefont {J.}~\bibnamefont {Theiler}},\ }\href@noop
  {} {\bibfield  {journal} {\bibinfo  {journal} {Journal of Neuroscience}\
  }\textbf {\bibinfo {volume} {18}},\ \bibinfo {pages} {7474} (\bibinfo {year}
  {1998})}\BibitemShut {NoStop}%
\bibitem [{\citenamefont {Visnovcova}\ \emph {et~al.}(2014)\citenamefont
  {Visnovcova}, \citenamefont {Mestanik}, \citenamefont {Javorka},
  \citenamefont {Mokra}, \citenamefont {Gala}, \citenamefont {Jurko},
  \citenamefont {Calkovska},\ and\ \citenamefont
  {Tonhajzerova}}]{visnovcova2014complexity}%
  \BibitemOpen
  \bibfield  {author} {\bibinfo {author} {\bibfnamefont {Z.}~\bibnamefont
  {Visnovcova}}, \bibinfo {author} {\bibfnamefont {M.}~\bibnamefont
  {Mestanik}}, \bibinfo {author} {\bibfnamefont {M.}~\bibnamefont {Javorka}},
  \bibinfo {author} {\bibfnamefont {D.}~\bibnamefont {Mokra}}, \bibinfo
  {author} {\bibfnamefont {M.}~\bibnamefont {Gala}}, \bibinfo {author}
  {\bibfnamefont {A.}~\bibnamefont {Jurko}}, \bibinfo {author} {\bibfnamefont
  {A.}~\bibnamefont {Calkovska}}, \ and\ \bibinfo {author} {\bibfnamefont
  {I.}~\bibnamefont {Tonhajzerova}},\ }\href@noop {} {\bibfield  {journal}
  {\bibinfo  {journal} {Physiological measurement}\ }\textbf {\bibinfo {volume}
  {35}},\ \bibinfo {pages} {1319} (\bibinfo {year} {2014})}\BibitemShut
  {NoStop}%
\bibitem [{\citenamefont {Yao}\ \emph {et~al.}(2018)\citenamefont {Yao},
  \citenamefont {Yao}, \citenamefont {Wu},\ and\ \citenamefont
  {Wang}}]{yao2018permutation}%
  \BibitemOpen
  \bibfield  {author} {\bibinfo {author} {\bibfnamefont {W.}~\bibnamefont
  {Yao}}, \bibinfo {author} {\bibfnamefont {W.}~\bibnamefont {Yao}}, \bibinfo
  {author} {\bibfnamefont {M.}~\bibnamefont {Wu}}, \ and\ \bibinfo {author}
  {\bibfnamefont {J.}~\bibnamefont {Wang}},\ }\href@noop {} {\bibfield
  {journal} {\bibinfo  {journal} {arXiv preprint arXiv:1801.05421}\ } (\bibinfo
  {year} {2018})}\BibitemShut {NoStop}%
\bibitem [{\citenamefont {Schindler}\ \emph {et~al.}(2016)\citenamefont
  {Schindler}, \citenamefont {Rummel}, \citenamefont {Andrzejak}, \citenamefont
  {Goodfellow}, \citenamefont {Zubler}, \citenamefont {Abela}, \citenamefont
  {Wiest}, \citenamefont {Pollo}, \citenamefont {Steimer},\ and\ \citenamefont
  {Gast}}]{schindler2016ictal}%
  \BibitemOpen
  \bibfield  {author} {\bibinfo {author} {\bibfnamefont {K.}~\bibnamefont
  {Schindler}}, \bibinfo {author} {\bibfnamefont {C.}~\bibnamefont {Rummel}},
  \bibinfo {author} {\bibfnamefont {R.~G.}\ \bibnamefont {Andrzejak}}, \bibinfo
  {author} {\bibfnamefont {M.}~\bibnamefont {Goodfellow}}, \bibinfo {author}
  {\bibfnamefont {F.}~\bibnamefont {Zubler}}, \bibinfo {author} {\bibfnamefont
  {E.}~\bibnamefont {Abela}}, \bibinfo {author} {\bibfnamefont
  {R.}~\bibnamefont {Wiest}}, \bibinfo {author} {\bibfnamefont
  {C.}~\bibnamefont {Pollo}}, \bibinfo {author} {\bibfnamefont
  {A.}~\bibnamefont {Steimer}}, \ and\ \bibinfo {author} {\bibfnamefont
  {H.}~\bibnamefont {Gast}},\ }\href@noop {} {\bibfield  {journal} {\bibinfo
  {journal} {Clinical neurophysiology}\ }\textbf {\bibinfo {volume} {127}},\
  \bibinfo {pages} {3051} (\bibinfo {year} {2016})}\BibitemShut {NoStop}%
\bibitem [{\citenamefont {Mart{\'\i}nez}\ \emph {et~al.}(2018)\citenamefont
  {Mart{\'\i}nez}, \citenamefont {Herrera-Diestra},\ and\ \citenamefont
  {Chavez}}]{martinez2018detection}%
  \BibitemOpen
  \bibfield  {author} {\bibinfo {author} {\bibfnamefont {J.~H.}\ \bibnamefont
  {Mart{\'\i}nez}}, \bibinfo {author} {\bibfnamefont {J.~L.}\ \bibnamefont
  {Herrera-Diestra}}, \ and\ \bibinfo {author} {\bibfnamefont {M.}~\bibnamefont
  {Chavez}},\ }\href@noop {} {\bibfield  {journal} {\bibinfo  {journal} {Chaos:
  An Interdisciplinary Journal of Nonlinear Science}\ }\textbf {\bibinfo
  {volume} {28}},\ \bibinfo {pages} {123111} (\bibinfo {year}
  {2018})}\BibitemShut {NoStop}%
\bibitem [{\citenamefont {Zanin}\ \emph {et~al.}(2018)\citenamefont {Zanin},
  \citenamefont {Rodr{\'\i}guez-Gonz{\'a}lez}, \citenamefont
  {Menasalvas~Ruiz},\ and\ \citenamefont {Papo}}]{zanin2018assessing}%
  \BibitemOpen
  \bibfield  {author} {\bibinfo {author} {\bibfnamefont {M.}~\bibnamefont
  {Zanin}}, \bibinfo {author} {\bibfnamefont {A.}~\bibnamefont
  {Rodr{\'\i}guez-Gonz{\'a}lez}}, \bibinfo {author} {\bibfnamefont
  {E.}~\bibnamefont {Menasalvas~Ruiz}}, \ and\ \bibinfo {author} {\bibfnamefont
  {D.}~\bibnamefont {Papo}},\ }\href@noop {} {\bibfield  {journal} {\bibinfo
  {journal} {Entropy}\ }\textbf {\bibinfo {volume} {20}},\ \bibinfo {pages}
  {665} (\bibinfo {year} {2018})}\BibitemShut {NoStop}%
\bibitem [{\citenamefont {Diks}\ \emph {et~al.}(1995)\citenamefont {Diks},
  \citenamefont {Van~Houwelingen}, \citenamefont {Takens},\ and\ \citenamefont
  {DeGoede}}]{diks1995reversibility}%
  \BibitemOpen
  \bibfield  {author} {\bibinfo {author} {\bibfnamefont {C.}~\bibnamefont
  {Diks}}, \bibinfo {author} {\bibfnamefont {J.}~\bibnamefont
  {Van~Houwelingen}}, \bibinfo {author} {\bibfnamefont {F.}~\bibnamefont
  {Takens}}, \ and\ \bibinfo {author} {\bibfnamefont {J.}~\bibnamefont
  {DeGoede}},\ }\href@noop {} {\bibfield  {journal} {\bibinfo  {journal}
  {Physics Letters A}\ }\textbf {\bibinfo {volume} {201}},\ \bibinfo {pages}
  {221} (\bibinfo {year} {1995})}\BibitemShut {NoStop}%
\bibitem [{\citenamefont {Daw}\ \emph {et~al.}(2000)\citenamefont {Daw},
  \citenamefont {Finney},\ and\ \citenamefont {Kennel}}]{daw2000symbolic}%
  \BibitemOpen
  \bibfield  {author} {\bibinfo {author} {\bibfnamefont {C.}~\bibnamefont
  {Daw}}, \bibinfo {author} {\bibfnamefont {C.}~\bibnamefont {Finney}}, \ and\
  \bibinfo {author} {\bibfnamefont {M.}~\bibnamefont {Kennel}},\ }\href@noop {}
  {\bibfield  {journal} {\bibinfo  {journal} {Physical Review E}\ }\textbf
  {\bibinfo {volume} {62}},\ \bibinfo {pages} {1912} (\bibinfo {year}
  {2000})}\BibitemShut {NoStop}%
\bibitem [{\citenamefont {Kennel}(2004)}]{kennel2004testing}%
  \BibitemOpen
  \bibfield  {author} {\bibinfo {author} {\bibfnamefont {M.~B.}\ \bibnamefont
  {Kennel}},\ }\href@noop {} {\bibfield  {journal} {\bibinfo  {journal}
  {Physical Review E}\ }\textbf {\bibinfo {volume} {69}},\ \bibinfo {pages}
  {056208} (\bibinfo {year} {2004})}\BibitemShut {NoStop}%
\bibitem [{\citenamefont {Costa}\ \emph {et~al.}(2008)\citenamefont {Costa},
  \citenamefont {Peng},\ and\ \citenamefont
  {Goldberger}}]{costa2008multiscale}%
  \BibitemOpen
  \bibfield  {author} {\bibinfo {author} {\bibfnamefont {M.~D.}\ \bibnamefont
  {Costa}}, \bibinfo {author} {\bibfnamefont {C.-K.}\ \bibnamefont {Peng}}, \
  and\ \bibinfo {author} {\bibfnamefont {A.~L.}\ \bibnamefont {Goldberger}},\
  }\href@noop {} {\bibfield  {journal} {\bibinfo  {journal} {Cardiovascular
  Engineering}\ }\textbf {\bibinfo {volume} {8}},\ \bibinfo {pages} {88}
  (\bibinfo {year} {2008})}\BibitemShut {NoStop}%
\bibitem [{\citenamefont {Casali}\ \emph {et~al.}(2008)\citenamefont {Casali},
  \citenamefont {Casali}, \citenamefont {Montano}, \citenamefont {Irigoyen},
  \citenamefont {Macagnan}, \citenamefont {Guzzetti},\ and\ \citenamefont
  {Porta}}]{casali2008multiple}%
  \BibitemOpen
  \bibfield  {author} {\bibinfo {author} {\bibfnamefont {K.~R.}\ \bibnamefont
  {Casali}}, \bibinfo {author} {\bibfnamefont {A.~G.}\ \bibnamefont {Casali}},
  \bibinfo {author} {\bibfnamefont {N.}~\bibnamefont {Montano}}, \bibinfo
  {author} {\bibfnamefont {M.~C.}\ \bibnamefont {Irigoyen}}, \bibinfo {author}
  {\bibfnamefont {F.}~\bibnamefont {Macagnan}}, \bibinfo {author}
  {\bibfnamefont {S.}~\bibnamefont {Guzzetti}}, \ and\ \bibinfo {author}
  {\bibfnamefont {A.}~\bibnamefont {Porta}},\ }\href@noop {} {\bibfield
  {journal} {\bibinfo  {journal} {Physical Review E}\ }\textbf {\bibinfo
  {volume} {77}},\ \bibinfo {pages} {066204} (\bibinfo {year}
  {2008})}\BibitemShut {NoStop}%
\bibitem [{\citenamefont {Donges}\ \emph {et~al.}(2013)\citenamefont {Donges},
  \citenamefont {Donner},\ and\ \citenamefont {Kurths}}]{donges2013testing}%
  \BibitemOpen
  \bibfield  {author} {\bibinfo {author} {\bibfnamefont {J.~F.}\ \bibnamefont
  {Donges}}, \bibinfo {author} {\bibfnamefont {R.~V.}\ \bibnamefont {Donner}},
  \ and\ \bibinfo {author} {\bibfnamefont {J.}~\bibnamefont {Kurths}},\
  }\href@noop {} {\bibfield  {journal} {\bibinfo  {journal} {EPL (Europhysics
  Letters)}\ }\textbf {\bibinfo {volume} {102}},\ \bibinfo {pages} {10004}
  (\bibinfo {year} {2013})}\BibitemShut {NoStop}%
\bibitem [{\citenamefont {Lacasa}\ \emph {et~al.}(2012)\citenamefont {Lacasa},
  \citenamefont {Nunez}, \citenamefont {Rold{\'a}n}, \citenamefont {Parrondo},\
  and\ \citenamefont {Luque}}]{lacasa2012time}%
  \BibitemOpen
  \bibfield  {author} {\bibinfo {author} {\bibfnamefont {L.}~\bibnamefont
  {Lacasa}}, \bibinfo {author} {\bibfnamefont {A.}~\bibnamefont {Nunez}},
  \bibinfo {author} {\bibfnamefont {{\'E}.}~\bibnamefont {Rold{\'a}n}},
  \bibinfo {author} {\bibfnamefont {J.~M.}\ \bibnamefont {Parrondo}}, \ and\
  \bibinfo {author} {\bibfnamefont {B.}~\bibnamefont {Luque}},\ }\href@noop {}
  {\bibfield  {journal} {\bibinfo  {journal} {The European Physical Journal B}\
  }\textbf {\bibinfo {volume} {85}},\ \bibinfo {pages} {217} (\bibinfo {year}
  {2012})}\BibitemShut {NoStop}%
\bibitem [{\citenamefont {Lacasa}\ and\ \citenamefont
  {Flanagan}(2015)}]{lacasa2015time}%
  \BibitemOpen
  \bibfield  {author} {\bibinfo {author} {\bibfnamefont {L.}~\bibnamefont
  {Lacasa}}\ and\ \bibinfo {author} {\bibfnamefont {R.}~\bibnamefont
  {Flanagan}},\ }\href@noop {} {\bibfield  {journal} {\bibinfo  {journal}
  {Physical Review E}\ }\textbf {\bibinfo {volume} {92}},\ \bibinfo {pages}
  {022817} (\bibinfo {year} {2015})}\BibitemShut {NoStop}%
\bibitem [{\citenamefont {Flanagan}\ and\ \citenamefont
  {Lacasa}(2016)}]{flanagan2016irreversibility}%
  \BibitemOpen
  \bibfield  {author} {\bibinfo {author} {\bibfnamefont {R.}~\bibnamefont
  {Flanagan}}\ and\ \bibinfo {author} {\bibfnamefont {L.}~\bibnamefont
  {Lacasa}},\ }\href@noop {} {\bibfield  {journal} {\bibinfo  {journal}
  {Physics Letters A}\ }\textbf {\bibinfo {volume} {380}},\ \bibinfo {pages}
  {1689} (\bibinfo {year} {2016})}\BibitemShut {NoStop}%
\bibitem [{\citenamefont {Graff}\ \emph {et~al.}(2013)\citenamefont {Graff},
  \citenamefont {Graff}, \citenamefont {Kaczkowska}, \citenamefont {Makowiec},
  \citenamefont {Amig{\'o}}, \citenamefont {Piskorski}, \citenamefont
  {Narkiewicz},\ and\ \citenamefont {Guzik}}]{graff2013ordinal}%
  \BibitemOpen
  \bibfield  {author} {\bibinfo {author} {\bibfnamefont {G.}~\bibnamefont
  {Graff}}, \bibinfo {author} {\bibfnamefont {B.}~\bibnamefont {Graff}},
  \bibinfo {author} {\bibfnamefont {A.}~\bibnamefont {Kaczkowska}}, \bibinfo
  {author} {\bibfnamefont {D.}~\bibnamefont {Makowiec}}, \bibinfo {author}
  {\bibfnamefont {J.}~\bibnamefont {Amig{\'o}}}, \bibinfo {author}
  {\bibfnamefont {J.}~\bibnamefont {Piskorski}}, \bibinfo {author}
  {\bibfnamefont {K.}~\bibnamefont {Narkiewicz}}, \ and\ \bibinfo {author}
  {\bibfnamefont {P.}~\bibnamefont {Guzik}},\ }\href@noop {} {\bibfield
  {journal} {\bibinfo  {journal} {The European Physical Journal Special
  Topics}\ }\textbf {\bibinfo {volume} {222}},\ \bibinfo {pages} {525}
  (\bibinfo {year} {2013})}\BibitemShut {NoStop}%
\bibitem [{\citenamefont {Bandt}\ and\ \citenamefont
  {Pompe}(2002)}]{bandt2002permutation}%
  \BibitemOpen
  \bibfield  {author} {\bibinfo {author} {\bibfnamefont {C.}~\bibnamefont
  {Bandt}}\ and\ \bibinfo {author} {\bibfnamefont {B.}~\bibnamefont {Pompe}},\
  }\href@noop {} {\bibfield  {journal} {\bibinfo  {journal} {Physical review
  letters}\ }\textbf {\bibinfo {volume} {88}},\ \bibinfo {pages} {174102}
  (\bibinfo {year} {2002})}\BibitemShut {NoStop}%
\bibitem [{\citenamefont {Zanin}\ \emph {et~al.}(2012)\citenamefont {Zanin},
  \citenamefont {Zunino}, \citenamefont {Rosso},\ and\ \citenamefont
  {Papo}}]{zanin2012permutation}%
  \BibitemOpen
  \bibfield  {author} {\bibinfo {author} {\bibfnamefont {M.}~\bibnamefont
  {Zanin}}, \bibinfo {author} {\bibfnamefont {L.}~\bibnamefont {Zunino}},
  \bibinfo {author} {\bibfnamefont {O.~A.}\ \bibnamefont {Rosso}}, \ and\
  \bibinfo {author} {\bibfnamefont {D.}~\bibnamefont {Papo}},\ }\href@noop {}
  {\bibfield  {journal} {\bibinfo  {journal} {Entropy}\ }\textbf {\bibinfo
  {volume} {14}},\ \bibinfo {pages} {1553} (\bibinfo {year}
  {2012})}\BibitemShut {NoStop}%
\bibitem [{\citenamefont {Schreiber}\ and\ \citenamefont
  {Schmitz}(1996)}]{schreiber1996improved}%
  \BibitemOpen
  \bibfield  {author} {\bibinfo {author} {\bibfnamefont {T.}~\bibnamefont
  {Schreiber}}\ and\ \bibinfo {author} {\bibfnamefont {A.}~\bibnamefont
  {Schmitz}},\ }\href@noop {} {\bibfield  {journal} {\bibinfo  {journal}
  {Physical Review Letters}\ }\textbf {\bibinfo {volume} {77}},\ \bibinfo
  {pages} {635} (\bibinfo {year} {1996})}\BibitemShut {NoStop}%
\bibitem [{\citenamefont {Schalk}\ \emph {et~al.}(2004)\citenamefont {Schalk},
  \citenamefont {McFarland}, \citenamefont {Hinterberger}, \citenamefont
  {Birbaumer},\ and\ \citenamefont {Wolpaw}}]{schalk2004bci2000}%
  \BibitemOpen
  \bibfield  {author} {\bibinfo {author} {\bibfnamefont {G.}~\bibnamefont
  {Schalk}}, \bibinfo {author} {\bibfnamefont {D.~J.}\ \bibnamefont
  {McFarland}}, \bibinfo {author} {\bibfnamefont {T.}~\bibnamefont
  {Hinterberger}}, \bibinfo {author} {\bibfnamefont {N.}~\bibnamefont
  {Birbaumer}}, \ and\ \bibinfo {author} {\bibfnamefont {J.~R.}\ \bibnamefont
  {Wolpaw}},\ }\href@noop {} {\bibfield  {journal} {\bibinfo  {journal} {IEEE
  Transactions on biomedical engineering}\ }\textbf {\bibinfo {volume} {51}},\
  \bibinfo {pages} {1034} (\bibinfo {year} {2004})}\BibitemShut {NoStop}%
\bibitem [{\citenamefont {Goldberger}\ \emph {et~al.}(2000)\citenamefont
  {Goldberger}, \citenamefont {Amaral}, \citenamefont {Glass}, \citenamefont
  {Hausdorff}, \citenamefont {Ivanov}, \citenamefont {Mark}, \citenamefont
  {Mietus}, \citenamefont {Moody}, \citenamefont {Peng},\ and\ \citenamefont
  {Stanley}}]{goldberger2000physiobank}%
  \BibitemOpen
  \bibfield  {author} {\bibinfo {author} {\bibfnamefont {A.~L.}\ \bibnamefont
  {Goldberger}}, \bibinfo {author} {\bibfnamefont {L.~A.}\ \bibnamefont
  {Amaral}}, \bibinfo {author} {\bibfnamefont {L.}~\bibnamefont {Glass}},
  \bibinfo {author} {\bibfnamefont {J.~M.}\ \bibnamefont {Hausdorff}}, \bibinfo
  {author} {\bibfnamefont {P.~C.}\ \bibnamefont {Ivanov}}, \bibinfo {author}
  {\bibfnamefont {R.~G.}\ \bibnamefont {Mark}}, \bibinfo {author}
  {\bibfnamefont {J.~E.}\ \bibnamefont {Mietus}}, \bibinfo {author}
  {\bibfnamefont {G.~B.}\ \bibnamefont {Moody}}, \bibinfo {author}
  {\bibfnamefont {C.-K.}\ \bibnamefont {Peng}}, \ and\ \bibinfo {author}
  {\bibfnamefont {H.~E.}\ \bibnamefont {Stanley}},\ }\href@noop {} {\bibfield
  {journal} {\bibinfo  {journal} {Circulation}\ }\textbf {\bibinfo {volume}
  {101}},\ \bibinfo {pages} {e215} (\bibinfo {year} {2000})}\BibitemShut
  {NoStop}%
\bibitem [{\citenamefont {Daniel}\ and\ \citenamefont
  {Lees}(1993)}]{daniel1993parkinson}%
  \BibitemOpen
  \bibfield  {author} {\bibinfo {author} {\bibfnamefont {S.}~\bibnamefont
  {Daniel}}\ and\ \bibinfo {author} {\bibfnamefont {A.}~\bibnamefont {Lees}},\
  }\href@noop {} {\bibfield  {journal} {\bibinfo  {journal} {Journal of neural
  transmission. Supplementum}\ }\textbf {\bibinfo {volume} {39}},\ \bibinfo
  {pages} {165} (\bibinfo {year} {1993})}\BibitemShut {NoStop}%
\bibitem [{\citenamefont {Lang}\ and\ \citenamefont
  {S}(1989)}]{lang1989assessment}%
  \BibitemOpen
  \bibfield  {author} {\bibinfo {author} {\bibfnamefont {A.~E.~T.}\
  \bibnamefont {Lang}}\ and\ \bibinfo {author} {\bibfnamefont {F.}~\bibnamefont
  {S}},\ }in\ \href@noop {} {\emph {\bibinfo {booktitle} {Quantification of
  neurological deficit}}}\ (\bibinfo  {publisher} {Butterworths},\ \bibinfo
  {year} {1989})\ pp.\ \bibinfo {pages} {285--309}\BibitemShut {NoStop}%
\bibitem [{\citenamefont {Hoehn}\ and\ \citenamefont
  {Yahr}(1967)}]{hoehn1967parkinsonism}%
  \BibitemOpen
  \bibfield  {author} {\bibinfo {author} {\bibfnamefont {M.~M.}\ \bibnamefont
  {Hoehn}}\ and\ \bibinfo {author} {\bibfnamefont {M.~D.}\ \bibnamefont
  {Yahr}},\ }\href@noop {} {\bibfield  {journal} {\bibinfo  {journal}
  {Neurology}\ }\textbf {\bibinfo {volume} {17}},\ \bibinfo {pages} {427}
  (\bibinfo {year} {1967})}\BibitemShut {NoStop}%
\bibitem [{\citenamefont {Shoeb}(2009)}]{shoeb2009application}%
  \BibitemOpen
  \bibfield  {author} {\bibinfo {author} {\bibfnamefont {A.~H.}\ \bibnamefont
  {Shoeb}},\ }\emph {\bibinfo {title} {Application of machine learning to
  epileptic seizure onset detection and treatment}},\ \href@noop {} {Ph.D.
  thesis},\ \bibinfo  {school} {Massachusetts Institute of Technology}
  (\bibinfo {year} {2009})\BibitemShut {NoStop}%
\bibitem [{\citenamefont {Olejarczyk}\ and\ \citenamefont
  {Jernajczyk}(2017)}]{olejarczyk2017graph}%
  \BibitemOpen
  \bibfield  {author} {\bibinfo {author} {\bibfnamefont {E.}~\bibnamefont
  {Olejarczyk}}\ and\ \bibinfo {author} {\bibfnamefont {W.}~\bibnamefont
  {Jernajczyk}},\ }\href@noop {} {\bibfield  {journal} {\bibinfo  {journal}
  {PloS one}\ }\textbf {\bibinfo {volume} {12}},\ \bibinfo {pages} {e0188629}
  (\bibinfo {year} {2017})}\BibitemShut {NoStop}%
\bibitem [{\citenamefont {Davidsdottir}\ \emph {et~al.}(2005)\citenamefont
  {Davidsdottir}, \citenamefont {Cronin-Golomb},\ and\ \citenamefont
  {Lee}}]{davidsdottir2005visual}%
  \BibitemOpen
  \bibfield  {author} {\bibinfo {author} {\bibfnamefont {S.}~\bibnamefont
  {Davidsdottir}}, \bibinfo {author} {\bibfnamefont {A.}~\bibnamefont
  {Cronin-Golomb}}, \ and\ \bibinfo {author} {\bibfnamefont {A.}~\bibnamefont
  {Lee}},\ }\href@noop {} {\bibfield  {journal} {\bibinfo  {journal} {Vision
  research}\ }\textbf {\bibinfo {volume} {45}},\ \bibinfo {pages} {1285}
  (\bibinfo {year} {2005})}\BibitemShut {NoStop}%
\bibitem [{\citenamefont {Shine}\ \emph {et~al.}(2011)\citenamefont {Shine},
  \citenamefont {Halliday}, \citenamefont {Naismith},\ and\ \citenamefont
  {Lewis}}]{shine2011visual}%
  \BibitemOpen
  \bibfield  {author} {\bibinfo {author} {\bibfnamefont {J.~M.}\ \bibnamefont
  {Shine}}, \bibinfo {author} {\bibfnamefont {G.~M.}\ \bibnamefont {Halliday}},
  \bibinfo {author} {\bibfnamefont {S.~L.}\ \bibnamefont {Naismith}}, \ and\
  \bibinfo {author} {\bibfnamefont {S.~J.}\ \bibnamefont {Lewis}},\ }\href@noop
  {} {\bibfield  {journal} {\bibinfo  {journal} {Movement Disorders}\ }\textbf
  {\bibinfo {volume} {26}},\ \bibinfo {pages} {2154} (\bibinfo {year}
  {2011})}\BibitemShut {NoStop}%
\bibitem [{\citenamefont {Papo}(2013{\natexlab{b}})}]{papo2013time}%
  \BibitemOpen
  \bibfield  {author} {\bibinfo {author} {\bibfnamefont {D.}~\bibnamefont
  {Papo}},\ }\href@noop {} {\bibfield  {journal} {\bibinfo  {journal}
  {Frontiers in physiology}\ }\textbf {\bibinfo {volume} {4}},\ \bibinfo
  {pages} {86} (\bibinfo {year} {2013}{\natexlab{b}})}\BibitemShut {NoStop}%
\bibitem [{\citenamefont {Martin}\ \emph {et~al.}(2001)\citenamefont {Martin},
  \citenamefont {Hudspeth},\ and\ \citenamefont
  {J{\"u}licher}}]{martin2001comparison}%
  \BibitemOpen
  \bibfield  {author} {\bibinfo {author} {\bibfnamefont {P.}~\bibnamefont
  {Martin}}, \bibinfo {author} {\bibfnamefont {A.}~\bibnamefont {Hudspeth}}, \
  and\ \bibinfo {author} {\bibfnamefont {F.}~\bibnamefont {J{\"u}licher}},\
  }\href@noop {} {\bibfield  {journal} {\bibinfo  {journal} {Proceedings of the
  National Academy of Sciences}\ }\textbf {\bibinfo {volume} {98}},\ \bibinfo
  {pages} {14380} (\bibinfo {year} {2001})}\BibitemShut {NoStop}%
\bibitem [{\citenamefont {Rupprecht}\ and\ \citenamefont
  {Prost}(2016)}]{rupprecht2016fresh}%
  \BibitemOpen
  \bibfield  {author} {\bibinfo {author} {\bibfnamefont {J.-F.}\ \bibnamefont
  {Rupprecht}}\ and\ \bibinfo {author} {\bibfnamefont {J.}~\bibnamefont
  {Prost}},\ }\href@noop {} {\bibfield  {journal} {\bibinfo  {journal}
  {Science}\ }\textbf {\bibinfo {volume} {352}},\ \bibinfo {pages} {514}
  (\bibinfo {year} {2016})}\BibitemShut {NoStop}%
\bibitem [{\citenamefont {Egolf}(2000)}]{egolf2000equilibrium}%
  \BibitemOpen
  \bibfield  {author} {\bibinfo {author} {\bibfnamefont {D.~A.}\ \bibnamefont
  {Egolf}},\ }\href@noop {} {\bibfield  {journal} {\bibinfo  {journal}
  {Science}\ }\textbf {\bibinfo {volume} {287}},\ \bibinfo {pages} {101}
  (\bibinfo {year} {2000})}\BibitemShut {NoStop}%
\bibitem [{\citenamefont {Lee}\ \emph {et~al.}(2001)\citenamefont {Lee},
  \citenamefont {Williams}, \citenamefont {Haig}, \citenamefont {Goldberg},\
  and\ \citenamefont {Gordon}}]{lee2001integration}%
  \BibitemOpen
  \bibfield  {author} {\bibinfo {author} {\bibfnamefont {K.-H.}\ \bibnamefont
  {Lee}}, \bibinfo {author} {\bibfnamefont {L.~M.}\ \bibnamefont {Williams}},
  \bibinfo {author} {\bibfnamefont {A.}~\bibnamefont {Haig}}, \bibinfo {author}
  {\bibfnamefont {E.}~\bibnamefont {Goldberg}}, \ and\ \bibinfo {author}
  {\bibfnamefont {E.}~\bibnamefont {Gordon}},\ }\href@noop {} {\bibfield
  {journal} {\bibinfo  {journal} {Clinical Neurophysiology}\ }\textbf {\bibinfo
  {volume} {112}},\ \bibinfo {pages} {1499} (\bibinfo {year}
  {2001})}\BibitemShut {NoStop}%
\bibitem [{\citenamefont {Ba{\c{s}}ar}\ and\ \citenamefont
  {G{\"u}ntekin}(2008)}]{bacsar2008review}%
  \BibitemOpen
  \bibfield  {author} {\bibinfo {author} {\bibfnamefont {E.}~\bibnamefont
  {Ba{\c{s}}ar}}\ and\ \bibinfo {author} {\bibfnamefont {B.}~\bibnamefont
  {G{\"u}ntekin}},\ }\href@noop {} {\bibfield  {journal} {\bibinfo  {journal}
  {Brain research}\ }\textbf {\bibinfo {volume} {1235}},\ \bibinfo {pages}
  {172} (\bibinfo {year} {2008})}\BibitemShut {NoStop}%
\bibitem [{\citenamefont {Roach}\ \emph {et~al.}(2013)\citenamefont {Roach},
  \citenamefont {Ford}, \citenamefont {Hoffman},\ and\ \citenamefont
  {Mathalon}}]{roach2013converging}%
  \BibitemOpen
  \bibfield  {author} {\bibinfo {author} {\bibfnamefont {B.}~\bibnamefont
  {Roach}}, \bibinfo {author} {\bibfnamefont {J.}~\bibnamefont {Ford}},
  \bibinfo {author} {\bibfnamefont {R.}~\bibnamefont {Hoffman}}, \ and\
  \bibinfo {author} {\bibfnamefont {D.}~\bibnamefont {Mathalon}},\ }in\
  \href@noop {} {\emph {\bibinfo {booktitle} {Supplements to Clinical
  neurophysiology}}},\ Vol.~\bibinfo {volume} {62}\ (\bibinfo  {publisher}
  {Elsevier},\ \bibinfo {year} {2013})\ pp.\ \bibinfo {pages}
  {163--180}\BibitemShut {NoStop}%
\bibitem [{\citenamefont {Oswal}\ \emph {et~al.}(2013)\citenamefont {Oswal},
  \citenamefont {Brown},\ and\ \citenamefont {Litvak}}]{oswal2013synchronized}%
  \BibitemOpen
  \bibfield  {author} {\bibinfo {author} {\bibfnamefont {A.}~\bibnamefont
  {Oswal}}, \bibinfo {author} {\bibfnamefont {P.}~\bibnamefont {Brown}}, \ and\
  \bibinfo {author} {\bibfnamefont {V.}~\bibnamefont {Litvak}},\ }\href@noop {}
  {\bibfield  {journal} {\bibinfo  {journal} {Current opinion in neurology}\
  }\textbf {\bibinfo {volume} {26}},\ \bibinfo {pages} {662} (\bibinfo {year}
  {2013})}\BibitemShut {NoStop}%
\bibitem [{\citenamefont {Little}\ and\ \citenamefont
  {Brown}(2014)}]{little2014functional}%
  \BibitemOpen
  \bibfield  {author} {\bibinfo {author} {\bibfnamefont {S.}~\bibnamefont
  {Little}}\ and\ \bibinfo {author} {\bibfnamefont {P.}~\bibnamefont {Brown}},\
  }\href@noop {} {\bibfield  {journal} {\bibinfo  {journal} {Parkinsonism \&
  related disorders}\ }\textbf {\bibinfo {volume} {20}},\ \bibinfo {pages}
  {S44} (\bibinfo {year} {2014})}\BibitemShut {NoStop}%
\bibitem [{\citenamefont {Rombouts}\ \emph {et~al.}(1995)\citenamefont
  {Rombouts}, \citenamefont {Keunen},\ and\ \citenamefont
  {Stam}}]{rombouts1995investigation}%
  \BibitemOpen
  \bibfield  {author} {\bibinfo {author} {\bibfnamefont {S.}~\bibnamefont
  {Rombouts}}, \bibinfo {author} {\bibfnamefont {R.}~\bibnamefont {Keunen}}, \
  and\ \bibinfo {author} {\bibfnamefont {C.}~\bibnamefont {Stam}},\ }\href@noop
  {} {\bibfield  {journal} {\bibinfo  {journal} {Physics Letters A}\ }\textbf
  {\bibinfo {volume} {202}},\ \bibinfo {pages} {352} (\bibinfo {year}
  {1995})}\BibitemShut {NoStop}%
\bibitem [{\citenamefont {Pezard}\ \emph {et~al.}(1994)\citenamefont {Pezard},
  \citenamefont {Martinerie}, \citenamefont {Breton}, \citenamefont
  {Bourzeix},\ and\ \citenamefont {Renault}}]{pezard1994non}%
  \BibitemOpen
  \bibfield  {author} {\bibinfo {author} {\bibfnamefont {L.}~\bibnamefont
  {Pezard}}, \bibinfo {author} {\bibfnamefont {J.}~\bibnamefont {Martinerie}},
  \bibinfo {author} {\bibfnamefont {F.}~\bibnamefont {Breton}}, \bibinfo
  {author} {\bibfnamefont {J.-C.}\ \bibnamefont {Bourzeix}}, \ and\ \bibinfo
  {author} {\bibfnamefont {B.}~\bibnamefont {Renault}},\ }\href@noop {}
  {\bibfield  {journal} {\bibinfo  {journal} {Electroencephalography and
  clinical neurophysiology}\ }\textbf {\bibinfo {volume} {91}},\ \bibinfo
  {pages} {383} (\bibinfo {year} {1994})}\BibitemShut {NoStop}%
\bibitem [{\citenamefont {Breakspear}\ and\ \citenamefont
  {Terry}(2002)}]{breakspear2002detection}%
  \BibitemOpen
  \bibfield  {author} {\bibinfo {author} {\bibfnamefont {M.}~\bibnamefont
  {Breakspear}}\ and\ \bibinfo {author} {\bibfnamefont {J.}~\bibnamefont
  {Terry}},\ }\href@noop {} {\bibfield  {journal} {\bibinfo  {journal}
  {Clinical neurophysiology}\ }\textbf {\bibinfo {volume} {113}},\ \bibinfo
  {pages} {735} (\bibinfo {year} {2002})}\BibitemShut {NoStop}%
\bibitem [{\citenamefont {Pezard}\ \emph {et~al.}(2001)\citenamefont {Pezard},
  \citenamefont {Jech},\ and\ \citenamefont
  {R{\u{u}}{\v{z}}i{\v{c}}ka}}]{pezard2001investigation}%
  \BibitemOpen
  \bibfield  {author} {\bibinfo {author} {\bibfnamefont {L.}~\bibnamefont
  {Pezard}}, \bibinfo {author} {\bibfnamefont {R.}~\bibnamefont {Jech}}, \ and\
  \bibinfo {author} {\bibfnamefont {E.}~\bibnamefont
  {R{\u{u}}{\v{z}}i{\v{c}}ka}},\ }\href@noop {} {\bibfield  {journal} {\bibinfo
   {journal} {Clinical Neurophysiology}\ }\textbf {\bibinfo {volume} {112}},\
  \bibinfo {pages} {38} (\bibinfo {year} {2001})}\BibitemShut {NoStop}%
\bibitem [{\citenamefont {Blesa}\ \emph {et~al.}(2017)\citenamefont {Blesa},
  \citenamefont {Trigo-Damas}, \citenamefont {Dileone}, \citenamefont {del
  Rey}, \citenamefont {Hernandez},\ and\ \citenamefont
  {Obeso}}]{blesa2017compensatory}%
  \BibitemOpen
  \bibfield  {author} {\bibinfo {author} {\bibfnamefont {J.}~\bibnamefont
  {Blesa}}, \bibinfo {author} {\bibfnamefont {I.}~\bibnamefont {Trigo-Damas}},
  \bibinfo {author} {\bibfnamefont {M.}~\bibnamefont {Dileone}}, \bibinfo
  {author} {\bibfnamefont {N.~L.-G.}\ \bibnamefont {del Rey}}, \bibinfo
  {author} {\bibfnamefont {L.~F.}\ \bibnamefont {Hernandez}}, \ and\ \bibinfo
  {author} {\bibfnamefont {J.~A.}\ \bibnamefont {Obeso}},\ }\href@noop {}
  {\bibfield  {journal} {\bibinfo  {journal} {Experimental neurology}\ }\textbf
  {\bibinfo {volume} {298}},\ \bibinfo {pages} {148} (\bibinfo {year}
  {2017})}\BibitemShut {NoStop}%
\bibitem [{\citenamefont {Buiatti}\ \emph {et~al.}(2007)\citenamefont
  {Buiatti}, \citenamefont {Papo}, \citenamefont {Baudonni{\`e}re},\ and\
  \citenamefont {van Vreeswijk}}]{buiatti2007feedback}%
  \BibitemOpen
  \bibfield  {author} {\bibinfo {author} {\bibfnamefont {M.}~\bibnamefont
  {Buiatti}}, \bibinfo {author} {\bibfnamefont {D.}~\bibnamefont {Papo}},
  \bibinfo {author} {\bibfnamefont {P.-M.}\ \bibnamefont {Baudonni{\`e}re}}, \
  and\ \bibinfo {author} {\bibfnamefont {C.}~\bibnamefont {van Vreeswijk}},\
  }\href@noop {} {\bibfield  {journal} {\bibinfo  {journal} {Neuroscience}\
  }\textbf {\bibinfo {volume} {146}},\ \bibinfo {pages} {1400} (\bibinfo {year}
  {2007})}\BibitemShut {NoStop}%
\bibitem [{\citenamefont {Gaspard}(2015)}]{gaspard2015cycles}%
  \BibitemOpen
  \bibfield  {author} {\bibinfo {author} {\bibfnamefont {P.}~\bibnamefont
  {Gaspard}},\ }\href@noop {} {\bibfield  {journal} {\bibinfo  {journal}
  {Chaos: An Interdisciplinary Journal of Nonlinear Science}\ }\textbf
  {\bibinfo {volume} {25}},\ \bibinfo {pages} {097606} (\bibinfo {year}
  {2015})}\BibitemShut {NoStop}%
\end{thebibliography}%

\end{document}